\documentclass[preprint,5p,times,twocolumn]{elsarticle}
\biboptions{sort&compress}

\usepackage{amssymb}
\usepackage{lipsum}  
\usepackage{placeins}
\usepackage{subcaption}
\usepackage{overpic}
\usepackage{url}
\usepackage[pdftex, pdftitle={Article}, pdfauthor={Author},colorlinks = true,allcolors = blue]{hyperref}
\usepackage{soul}
\usepackage{calrsfs}
\DeclareMathAlphabet{\pazocal}{OMS}{zplm}{m}{n}

\newcommand{\Lb}{\pazocal{L}}
\usepackage{lineno}
\journal{NIMA}

\begin{document}


\title{Feasibility Study of Measuring $\Lambda^0\to n\pi^{0}$ Using a High-Granularity \\Zero-Degree Calorimeter at the Future Electron-Ion Collider}

\author[inst1]{Sebouh J. Paul}

\author[inst1]{Ryan Milton}
\author[inst1]{Sebasti\'an Mor\'an}
\author[inst1]{Barak Schmookler}
\author[inst1]{Miguel Arratia\fnref{fn1}}

\address[inst1]{Department of Physics and Astronomy, University of California, Riverside, CA 92521, USA}

\fntext[fn1]{corresponding author, miguel.arratia@ucr.edu}

\begin{abstract}
Key measurements at the future Electron-Ion Collider (EIC), including first-of-their-kind studies of kaon structure, require the detection of $\Lambda^0$ at forward angles. We present a feasibility study of $\Lambda^0 \rightarrow n\pi^0$ measurements using a high-granularity Zero Degree Calorimeter to be located about 35 m from the interaction point. We introduce a method to address the unprecedented challenge of identifying $\Lambda^0$s with energy $O(100)$ GeV that produce displaced vertices of $O(10)$ m. In addition, we present a reconstruction approach using graph neural networks. We find that the energy and angle resolution for $\Lambda^0$ is similar to that for neutrons, both of which meet the requirements outlined in the EIC Yellow Report. Furthermore, we estimate performance for measuring the neutron’s direction in the $\Lambda^0$ rest frame, which reflects the $\Lambda^0$ spin polarization. We estimate that the neutral-decay channel $\Lambda^0 \rightarrow n\pi^0$ will greatly extend the measurable energy range for the charged-decay channel $\Lambda^0 \rightarrow p\pi^-$, which is limited by the location of small-angle trackers and the accelerator magnets. This work paves the way for EIC studies of kaon structure and spin phenomena.
 
\end{abstract}

\maketitle
\section{Introduction}
\label{sec:outline}

The Electron-Ion Collider (EIC)~\cite{Accardi:2012qut} physics program will enable groundbreaking measurements of the structure of hadrons and atomic nuclei by facilitating the first-ever collisions of polarized electrons with polarized protons and light nuclei, as well as collisions involving polarized electrons and heavy nuclei. This program will be carried out using a detector system called ePIC, which consists of a main detector covering full azimuth and $|\eta| < 4$, along with several subdetectors positioned at forward angles, all closely integrated with the accelerator beamline~\cite{Adkins:2022jfp,ATHENA:2022hxb}. The forward detectors will play a key role in exclusive reactions that produce high-energy hadrons near beam rapidity, such as deep-exclusive meson production (DEMP).

For example, DEMP studies where a neutron is measured at small angles, $ep \to e \pi^{+} n$, will constrain pion form factors. Similarly, the reaction $ep \to e K^{+} \Lambda^{0}$ with forward $\Lambda^{0}$ will constrain kaon form factors. Combined, these studies could address questions such as: ``What is the size and range of interference between emergent mass and the Higgs mass mechanism?''~\cite{EICYR}. Additionally, a variety of channels, such as $ep \to e X \Lambda^0$--- where $X$ may include a jet or photon---will enable broader studies of the quark-gluon structure of kaons. A summary of the relevant reactions for pion and kaon structure and the physics they aim to explore is provided in Refs.~\cite{Horn:2016rip,Aguilar:2019teb,Arrington:2021biu,EICYR}. 

\begin{figure}[h!]
    \centering
    \includegraphics[width=\linewidth]{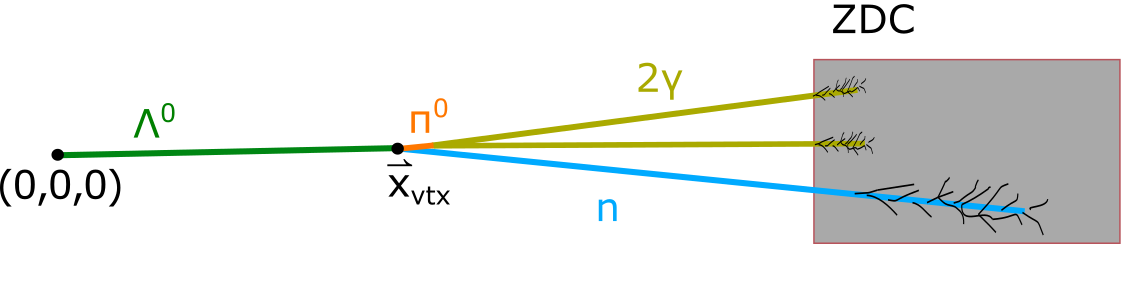}
    \caption{Topology of $\Lambda^0$ decay in neutral-channel decay.  Not shown to scale.}
    \label{fig:topology}
\end{figure}
While the feasibility of measuring the DEMP via the neutron channel with a Zero-Degree Calorimeter (ZDC) has been studied extensively~\cite{EICYR,Bylinkin:2022rxd,Arrington:2021biu}, the prospects of the $\Lambda^{0}$ channel remain much less understood. The $\Lambda^{0}$ could, in principle, be measured either in the charged-decay channel $\Lambda^{0}\rightarrow p\pi^-$ with trackers or in the neutral-decay channel $\Lambda^0 \to n \pi^{0}$ with a ZDC (as illustrated in Figure~\ref{fig:topology}), and either way, it is extremely challenging. The report on the science requirements and detector concepts for the EIC, commonly referred to as the EIC Yellow Report (YR)~\cite{EICYR}, states that ``{\it The reconstruction of the $\Lambda^{0}$ event in the far-forward detection area is one of the most challenging tasks. This comes mainly from the fact that these leading $\Lambda^0$'s have energy close to the initial beam energy, and thus their decay lengths can be tens of meters along the Z-axis (or beam-line). This complicates detection of the decay products, and thus the final $\Lambda^0$ mass reconstruction.}''

The feasibility of measuring $\Lambda^{0}\rightarrow p\pi^-$ (branching ratio 64\%~\cite{ParticleDataGroup:2024cfk}) using near-beamline trackers was explored in the EIC YR~\cite{EICYR} and Ref.~\cite{Tu:2023few}, as well as in studies~\cite{Ji:2023cdh} for the EicC in China~\cite{Anderle:2021wcy}. For the EIC, this charged-decay channel was found feasible only at the lowest-energy setting (electron beam energy of 5 GeV and proton beam energy of 41 GeV, abbreviated as 5 $\times$ 41 GeV), as higher-energy $\Lambda^0$s are less likely to decay before reaching these trackers, and the acceptance is constrained by the accelerator magnets.

In contrast, the prospects of measuring  $\Lambda^0 \to n \pi^{0}$ (branching ratio 36\%~\cite{ParticleDataGroup:2024cfk}) with the ZDC has not been studied with detailed simulations. For instance, the EIC YR~\cite{EICYR} only presents kinematic distributions for the decay products of $\Lambda^0 \to n\pi^0$ without detector simulations, from which only limited information can be inferred.  Arrington \textit{et al.} suggested from this kinematics study the possibility of measuring this decay channel combining the measurement of the neutron in the ZDC with the measurement of the photons from the $\pi^0$ decay in an electromagnetic calorimeter in the B0 region~\cite{Arrington:2021biu}. 
Similarly, Ref.~\cite{Xie:2021ypc} presents a similar study for the EicC. To date, no prior study has demonstrated feasibility taking into account detector response or clustering algorithms and particle identification algorithms.

In this work, we present the first feasibility study for $\Lambda^0 \to n\pi^0$. Specifically, we demonstrate that the high-granularity ZDC design proposed in Ref.~\cite{Milton:2024bqv}, along with the methods described here, successfully addresses this challenge. Section~\ref{sec:simulation} details our simulations, Section~\ref{sec:reconstruction} outlines the reconstruction strategy, Section~\ref{sec:performance} presents the performance results, and Section~\ref{sec:summary} provides a summary of the results and discussion.

\section{Simulation}
\label{sec:simulation}
\subsection{Kinematics of $\Lambda^0$ decay }
We generate a sample of 1 million events, each containing a single decay of $\Lambda^0 \to n\pi^{0}$ and no other particles. The simulated $\Lambda^0$ is assigned a mass of $1116$ MeV~\cite{ParticleDataGroup:2024cfk} and has energies uniformly distributed in log space between 50 GeV and 300 GeV, with uniformly distributed polar angles between 0 and 3 mrad and azimuthal angles between 0 and $2\pi$ with respect to the EIC's proton beam axis\footnote{The $\Lambda^0$ generated along the proton axis in $+z$---the hadron-going side of the experiment. The $+x$ direction is towards the inner part of the EIC ring, and $+y$ points upwards. The polar angle, $\theta$, is defined as the angle from the $+z$ axis. $\phi$, the azimuthal angle, is the angle from $+x$ in the $xy$ plane and increases counterclockwise. Note that the proton beam is angled 25 mrad relative to the electron beam.}. We do not include any beam effects in the simulation.

The distance traveled by the $\Lambda^0$ before decaying follows an exponential distribution with a mean value of $\beta\gamma c\tau$, where $\beta$ is the velocity of the $\Lambda^0$ as a fraction of the speed of light, $\gamma = 1/\sqrt{1-\beta^2}$ is the relativistic correction factor, $c$ is the speed of light, and $\tau$ is the mean lifetime of a $\Lambda^0$, $2.62\times 10^{-10}$~s~\cite{ParticleDataGroup:2024cfk}. 

In the simulation, the $\Lambda^0$ then decays isotropically in its rest frame into a neutron and a $\pi^0$, which promptly decays isotropically in its rest frame into two photons.  Consequently, the neutron carries between 75\% and 94\% of the energy of the $\Lambda^0$ in the lab frame, regardless of the momentum of the $\Lambda^0$.  This is because the neutron is much more massive than the $\pi^0$, and in the  $\Lambda^0$ center-of-mass frame, the momenta of both of these two particles is around 100~MeV, which is relatively small compared to the mass of the neutron.  

We also generated 500k $\Lambda^0$ events with each event assigned one of 20 unique discrete energy values sampled uniformly in log space between 50 and 300 GeV. The angular distributions are the same as the continuous data. We use the 1 million event continuous dataset and 500k event discrete dataset for training and testing the AI models in Section~\ref{sec:AI}, respectively. 

For studies comparing $\Lambda^0$ performance to neutron performance, we generated single neutron datasets with the same energy and angle distributions as the $\Lambda^0$ datasets. 

For all plots in this paper, the 500k event dataset with discrete energy values are shown.

\subsection{Detector Response}
Following Ref.~\cite{Milton:2024bqv}, the \textsc{Geant4}~\cite{GEANT4:2002zbu} (v11.02.p2) simulations of the ZDC geometry~\cite{Milton:2024bqv} were implemented\footnote{The ePIC ZDC design includes a removable crystal calorimeter for low-energy photons ($O(100)$ MeV), but this work focuses on the SiPM-on-tile component~\cite{Milton:2024bqv}, as the crystal calorimeter is unnecessary for $\Lambda^0$ reconstruction and would overly complicate the analysis.} in the \textsc{DD4HEP} framework~\cite{Frank:2014zya} using the FTFP\_BERT physics list. No noise was included in the simulation. This simulation framework was validated by comparing it to test-beam data~\cite{CALICE:2012eac,CALICE:2022uwn} and GNN studies~\cite{CALICE:2024imr} from the CALICE collaboration, as discussed in previous papers~\cite{Milton:2024bqv,Insert,Arratia:2023xhz,Arratia:2023rdo}.  

Following our previous studies~\cite{Milton:2024bqv,Insert}, hits in the ZDC are reconstructed using a 15-bit ADC with no noise and a dynamic range of 800 MeV. These hits are included in the subsequent event reconstruction only if they exceed an energy threshold of 0.5 MIPs\footnote{One MIP is the most-probable hit energy deposited by a minimum-ionizing particle, which was determined using muon simulations to be 0.47 MeV.}, and the hit time is less than 270 ns\footnote{that is, the amount of time for a particle moving at the speed of light to reach the ZDC front face from the interaction point, plus an additional 150 ns.}. These cuts were inspired by those used in CALICE prototypes~\cite{CALICE:2012eac,CALICE:2022uwn} and the anticipated capabilities of the EIC ASIC for SiPM readout. Hits passing this selection are simply referred as hits in the rest of this work.

\begin{figure*}[h!]
    \centering
    \includegraphics[width=0.33\linewidth]{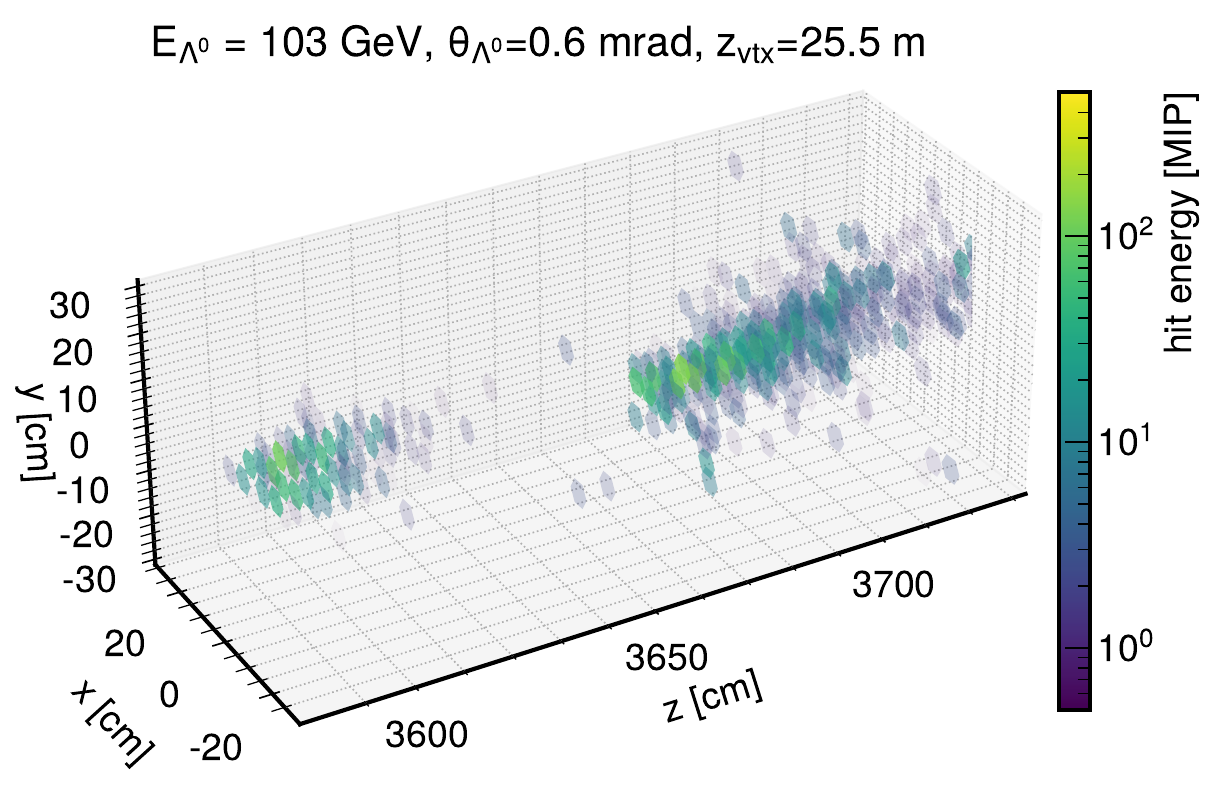}\includegraphics[width=0.33\linewidth]{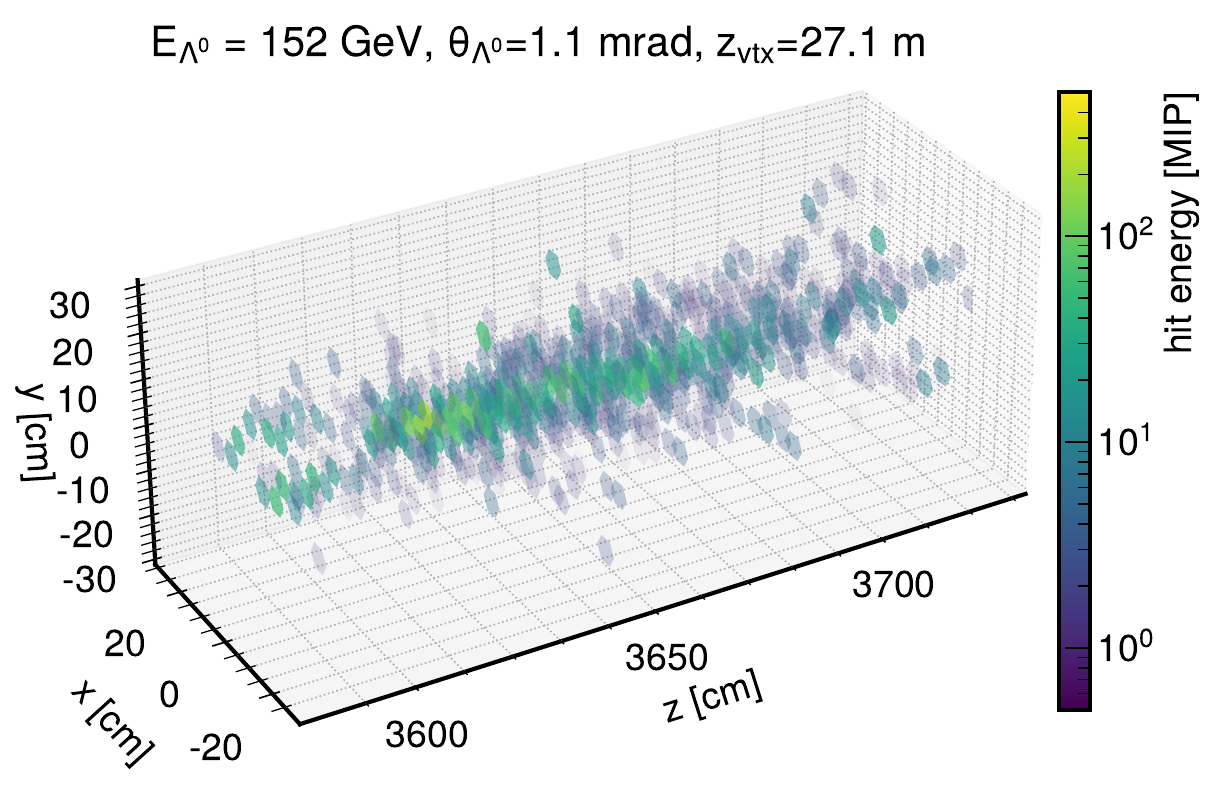}\includegraphics[width=0.33\linewidth]{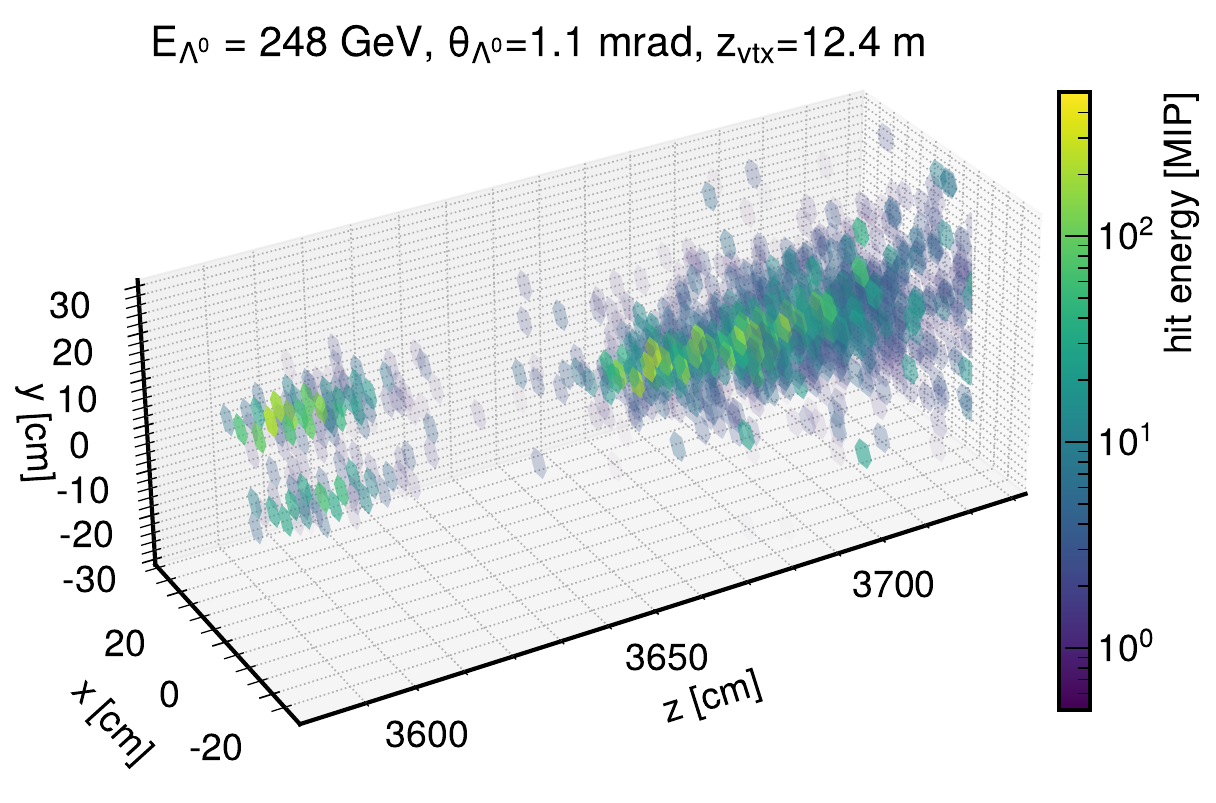}
    \includegraphics[width=0.33\linewidth]{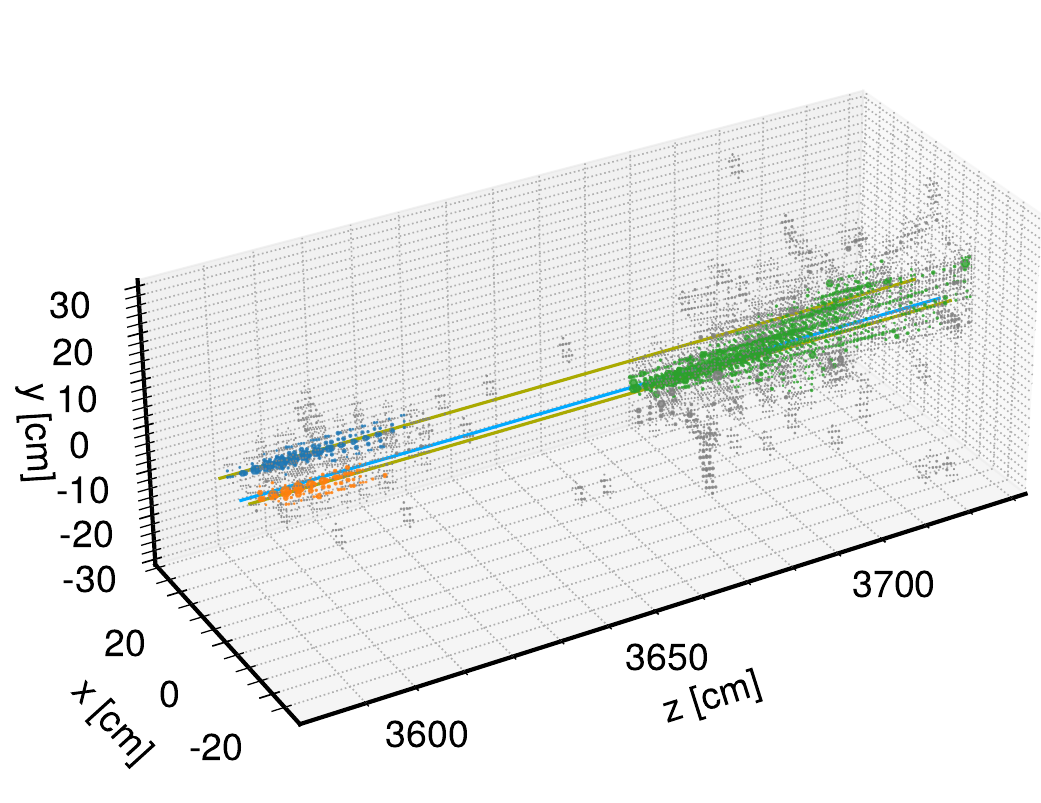}\includegraphics[width=0.33\linewidth]{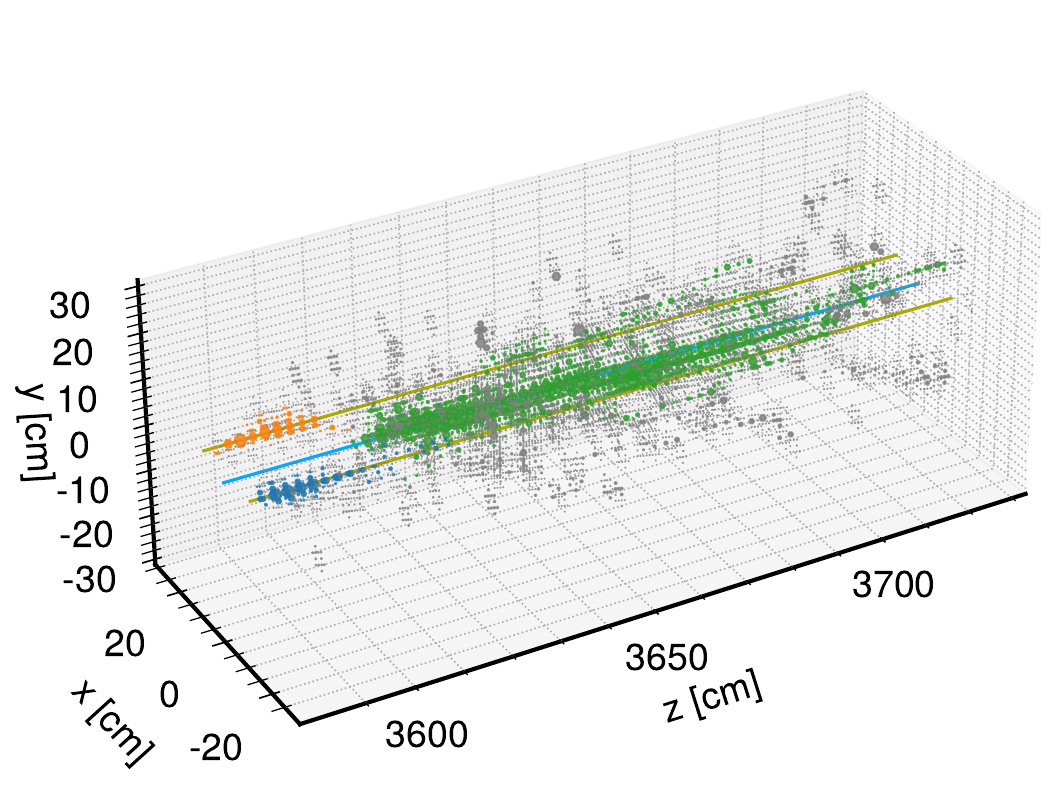}\includegraphics[width=0.33\linewidth]{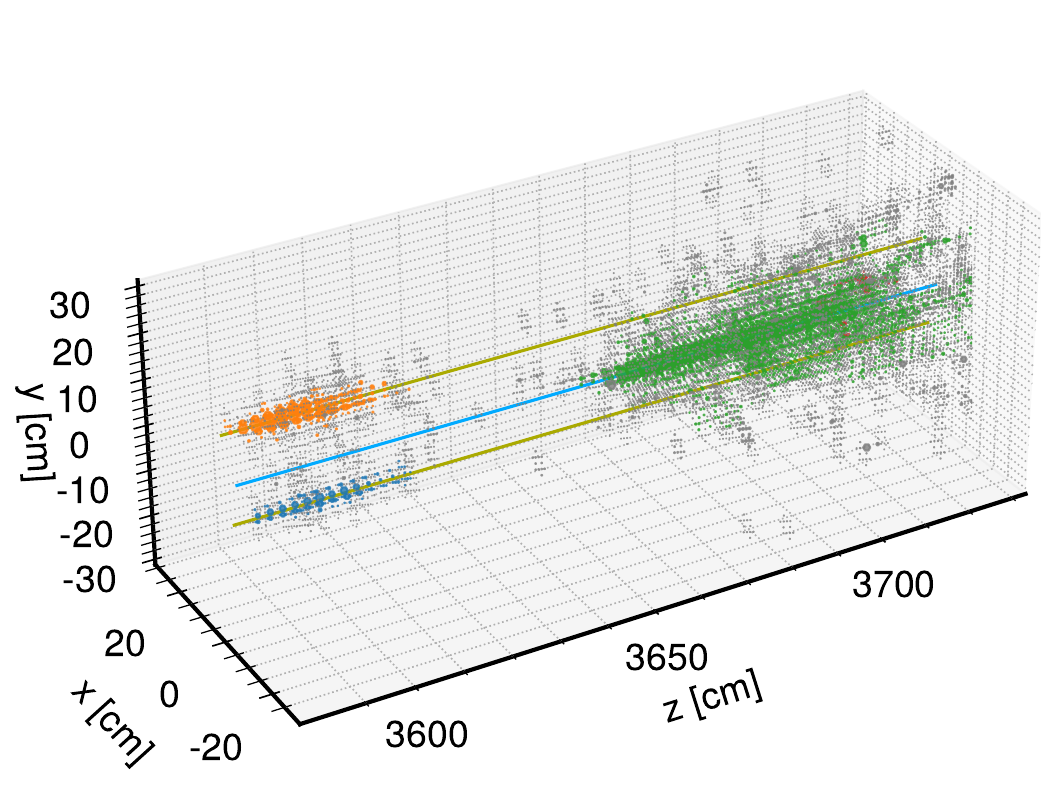}
    \caption{Top row: event display for example events, with the hits color coded by energy deposited, in units of MIPs.  Bottom row:  the same events, with subcell hits from HEXPLIT algorithm color-coded by which cluster the subcell hits have been assigned to by the 3D topological cluster described in the text.}
    \label{fig:event_display}
\end{figure*}

Example events are shown in Fig.~\ref{fig:event_display}. The events shown were chosen as good-quality events, where the particles were far enough apart that they produced distinguishable showers. 

\section{$\Lambda^{0}$ Reconstruction}
\label{sec:reconstruction}

\subsection{Geometric Acceptance}
The geometrical acceptance for $\Lambda^0 \rightarrow n\pi^0$ is defined as the probability that the $\Lambda^0$ decays before reaching the front face of the ZDC and that all three daughter particles strike the ZDC within its fiducial region\footnote{This fiducial region is determined by the aperture of the magnets and the design of the beam-pipe exit window before the ZDC.}, defined by a radius of 14.3 cm.  For comparison, the transverse size of the detector is $\approx60\times60$ cm$^2$.

The geometrical acceptance depends strongly on the $\Lambda^{0}$ energy, as it determines the boost that influences the opening angles between particles and the distribution of the displaced vertex position, $z_{\rm vtx}$, which defines the solid angle covered by the ZDC. 

\begin{figure*}
    \centering
    \includegraphics[width=\linewidth]{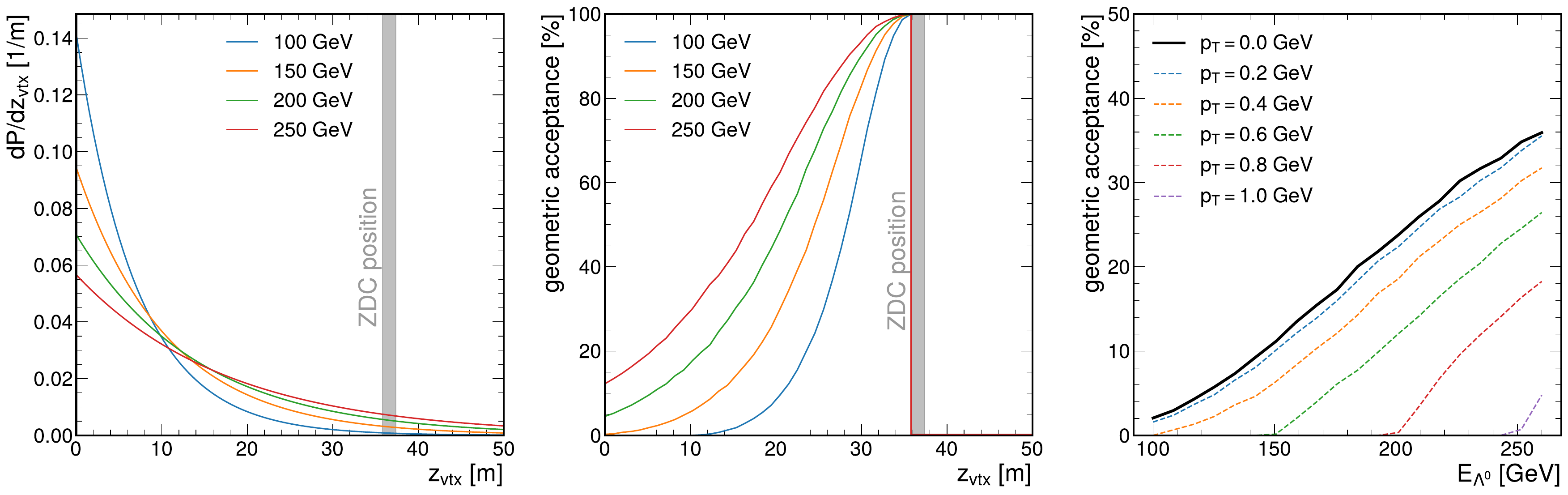}
    \caption{Left:  Decay position distribution, for different $\Lambda^0$ energies.  Middle: acceptance as a function of energy and decay position.  Right:  Acceptance as a function of energy, integrated over decay position, for $\Lambda^0$s along the proton axis (black solid curve) and with varying amounts of transverse momentum (colored dashed curves).}
    \label{fig:acceptance}
\end{figure*}

Figure~\ref{fig:acceptance}, left panel, shows the $z_{\rm vtx}$ distributions for different $\Lambda^0$ energies. The middle panel presents the geometrical acceptance as a function of energy and $z_{\rm vtx}$. We then determined the geometrical acceptance as a function of energy, integrated over $z_{\rm vtx}$.  We evaluated this for $\Lambda^0$s along the proton beam axis ($p_T=0$) and show this as a solid black line in the right panel of Fig.~\ref{fig:acceptance}.  We find that acceptance increases almost linearly from around 2\% at 100~GeV to around 35\% at 250~GeV.  For comparison, in the YR~\cite{EICYR} it was estimated that acceptance for detecting $\Lambda^{0}\rightarrow p\pi^-$ is on average around 1\% for the $18\times275$~GeV beam setting. We also evaluated the geometrical acceptance for off-axis $\Lambda^0$s, with $p_T$ up to 1 GeV, and show the results as dashed curves in the right panel of Fig.~\ref{fig:acceptance}.  The acceptance at $p_T=0.2$~GeV and 0.4~GeV is comparable to the acceptance at $p_T=0$~GeV, but the acceptance drops at higher $p_T$ values.

\subsection{Energy reconstruction and Clustering}
\label{sec:conventional_recon}
In order to improve the effective transverse granularity, and therefore obtain better spatial resolution on the incident particle positions, the HEXPLIT algorithm~\cite{hexplit}, summarized below, was used.  HEXPLIT infers the distribution of energy deposited within individual cells, taking advantage of the overlapping of the staggered layers of the ZDC.  This splits the hits in each cell into ``subcell" hits.  The ZDC design~\cite{Milton:2024bqv} uses a 4-layer staggering cycle, and therefore the redistribution of energy between subcells of a cell depends on the energy in the overlapping cells in layers immediately upstream and downstream of the current layer and those that are two layers upstream or downstream of the current cell.  We used the following formula
\begin{equation}
W_i = \prod\limits_{j=1}^{3}{\rm max}(E_{j},\delta),
\label{eq:Wi_analytical}
\end{equation}
Here, $i$ denotes the index of a subcell.  The product over $j$ is over the cells in the nearby layers that overlap with subcell $i$, and $E_j$ represents the energy in that cell.  The tolerance parameter $\delta$ prevents division by zero in the case that there are no overlapping hits on a neighboring layer, and is set to be much smaller than any real hit's energy (0.01~MIPs) in order to have negligible impact in all other cases.  Since the layouts on the layers that are two layers distant from the current layer exhibit identical patterns, we introduce a virtual cell, whose energy is the aggregate of the energies from the cells within those layers overlapping the $i^{\rm th}$ subcell.  

We proceed to assign the updated signal within the current subcell as follows:
\begin{equation}
E_i = E_{\rm cell} W_i/\sum_j W_j.
\end{equation}
Here, the summation of weights across $j$ serves to normalize the weights, ensuring that the total signal within the cell remains unchanged, \textit{i.e}., $\sum_i E_i = E_{\rm cell}$.

The topoclustering algorithm~\cite{ATLAS:2016krp} then determined which subcell hits belong to which clusters, as follows.  First, we filtered out all subcell hits with energy below 50 keV (equivalent to $\approx$0.1 MIPs). Then we determined which hits are neighbors to one another.  Pairs of hits were considered neighbors if they are in the same or an adjacent subcell position as one another and on the same or an adjacent layer to one another.  Any hits with energy greater than 3 MeV ($\approx$6.4~MIPs) were considered seed hits for defining a cluster.  All hits that are neighbors of the seed hits, and the neighbors of those hits, and so on, are then included in the cluster.  Clusters were then accepted if they have at least 30 subcell hits, and a total energy of at least 11 MeV ($\approx$23 MIPs).  

The results of the topoclustering algorithm applied to a selected sample of $\Lambda^0 \rightarrow n\pi^0$ events passing our fiducial selection are shown in the bottom row of Fig~\ref{fig:event_display}. 

\begin{figure}
    \centering
    \includegraphics[width=\linewidth]{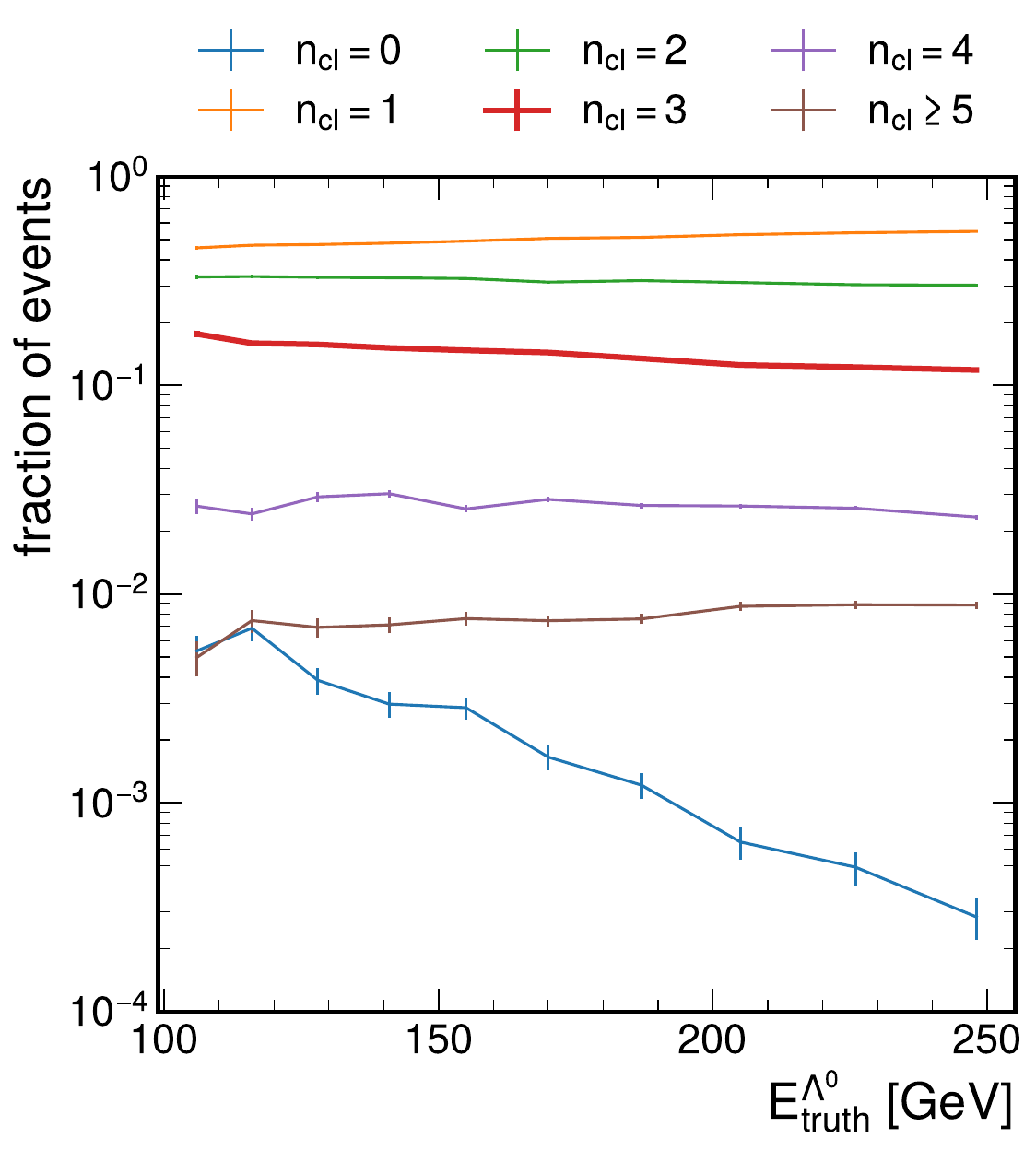}
    \caption{Number of clusters per $\Lambda^0$ event, as found by the topo-clustering algorithm, as a function of the $\Lambda^0$'s energy.}
    \label{fig:clust_count}
\end{figure}

Figure~\ref{fig:clust_count} shows the number of clusters found per event as a function of the $\Lambda^0$ energy. Depending on the energy, about 10--20$\%$ of the events passing the fiducial selection yield 3 topoclusters or more, which is sufficient, in principle, to reconstruct the $\Lambda^{0}$ invariant mass using conventional methods. Events with significant overlap between neutron and photon showers will be addressed using AI methods, as detailed in Sec.~\ref{sec:AI}.

We then determined the energy of each cluster at the electromagnetic scale as the sum of the energies of each subcell hit, divided by the sampling fraction of the detector ($SF$=2.03\%), which was defined using an electron simulation without including clustering effects.  

The position of the clusters were determined using a weighted average of the subcell hit positions,
\begin{equation}
    \vec x_{\rm cl}=\frac{\sum\limits_{i\in\rm hits}\vec x_i w_i}{\sum\limits_{i\in\rm hits} w_i}
\end{equation}
where each subcell hit's weight is given by
\begin{equation}
    w_i=\max(0, w_0+\log E_i/E_{\rm tot})
\end{equation}
where $w_0$ is an offset which determines the minimum energy fraction of subcell hits used in the position determination, and depends on the total energy of all subcell hits in the cluster, $E_{\rm tot}$, and $E_i$ is the energy of the subcell hit.  The value of $w_0$ used is given by the following parameterization from Ref.~\cite{hexplit}:
\begin{equation}
    w_0=5.8+0.65 u+0.31 u^2,
\end{equation}
where 
\begin{equation}
    u=\log \frac{E_{\rm tot}/SF}{50\,\,\rm GeV},
\end{equation}
yielding values of $w_0$ between 5.9 and 7.3 for the energy range of clusters in this analysis.

We also determined the moment matrix of each cluster, given by 
\begin{equation}
    \textbf M =\frac{\sum\limits_{i\in\rm hits} w_i(\vec x_i-\vec x_{\rm cl})\otimes(\vec x_i-\vec x_{\rm cl})}{\sum\limits_{i\in\rm hits} w_i}.
\end{equation}

\begin{figure*}
    \centering
    \includegraphics[width=.32\linewidth]{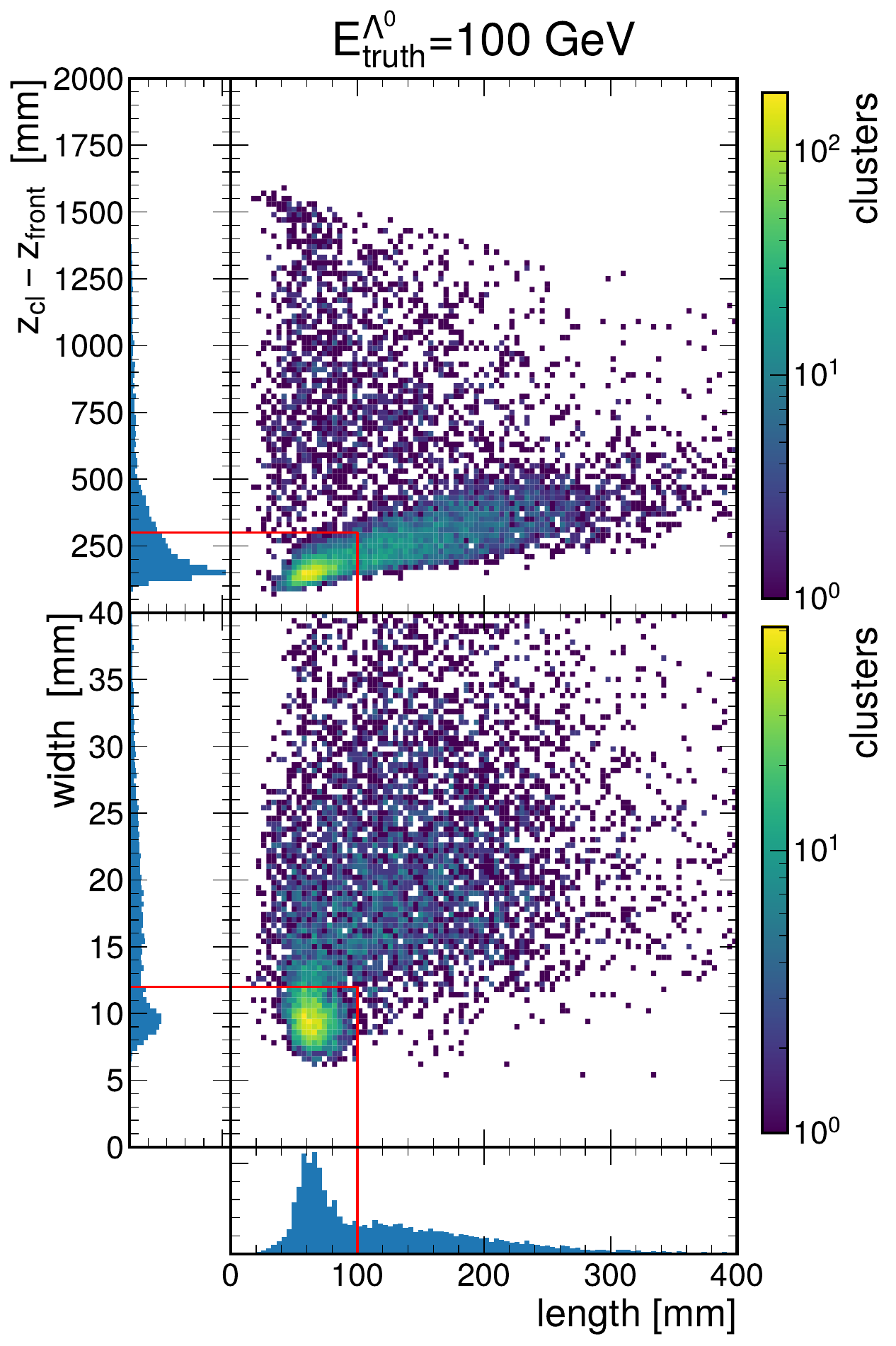}\includegraphics[width=.32\linewidth]{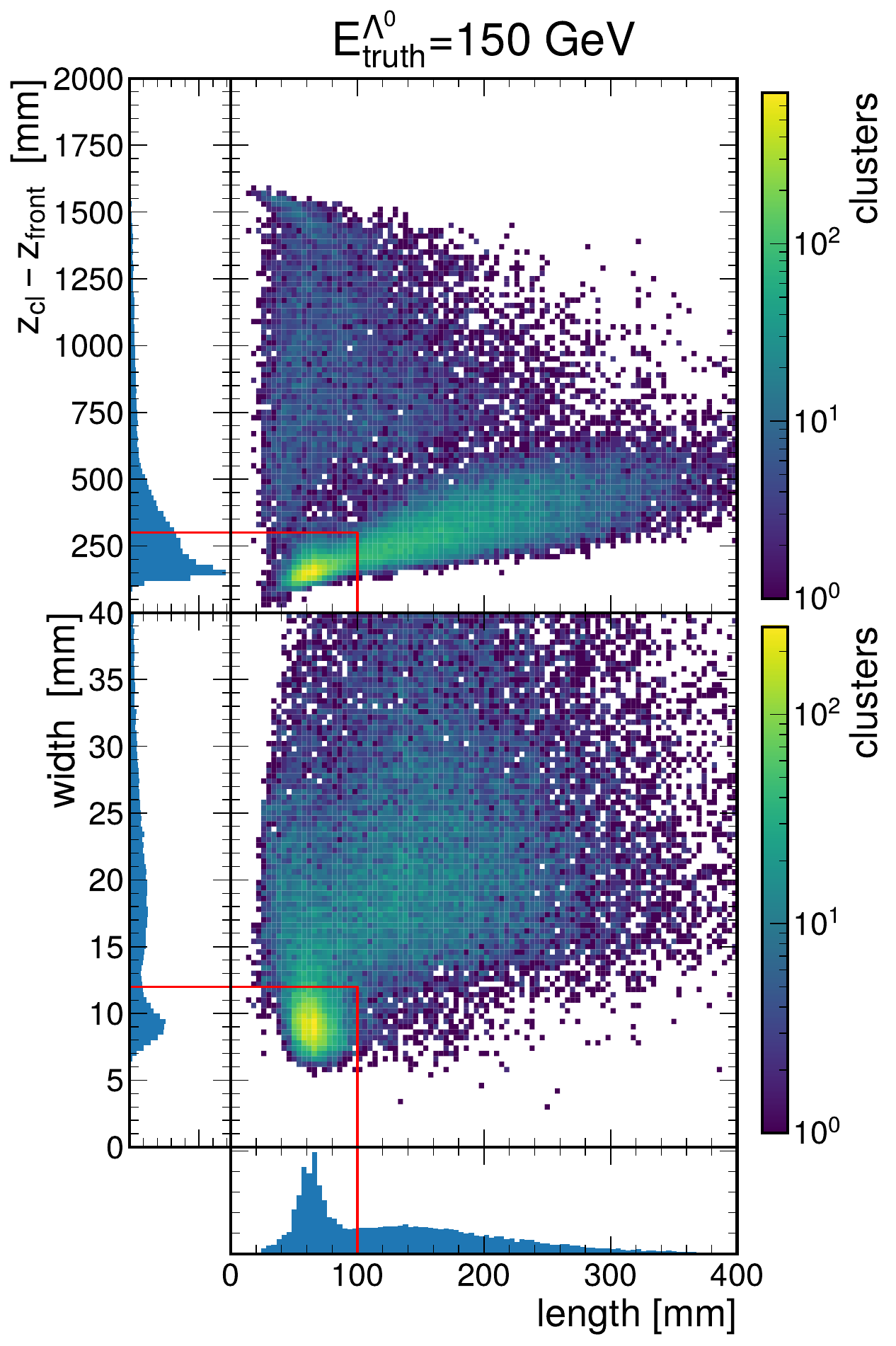}\includegraphics[width=.32\linewidth]{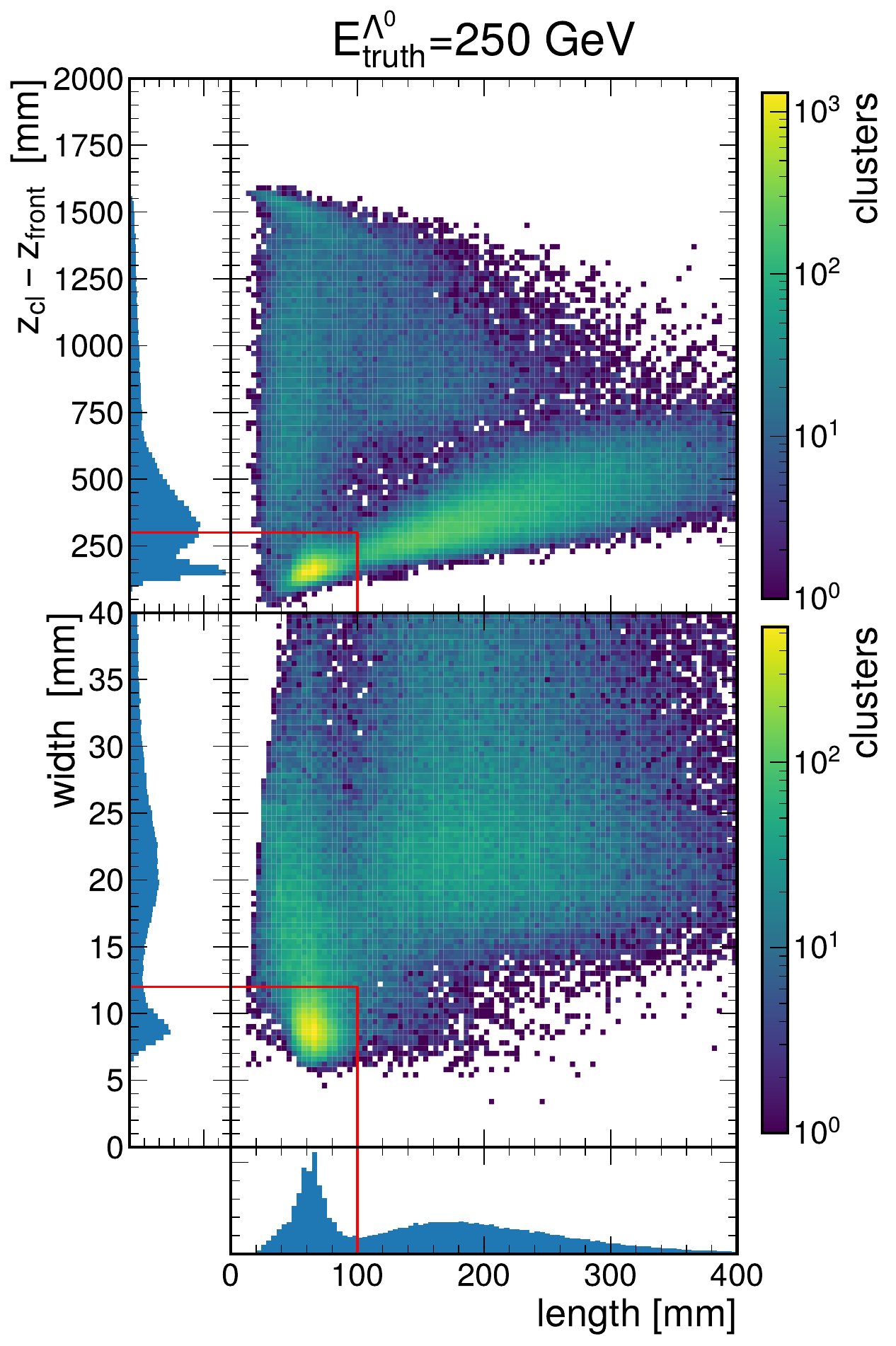}
    \caption{Distribution of the cluster shape parameters, at three different $\Lambda^0$ energies.  The red lines show the cuts used to identify the photon clusters.}
    \label{fig:cluster_shape}
\end{figure*}

To identify clusters associated with photons, we looked for clusters with a short length, $\ell$, (defined as the square root of the largest eigenvalue in the moment matrix) small transverse width, $w_{T}$, (defined as the square root of the second largest eigenvalue), and which had positions close to the front face of the detector.  We show the distributions of these parameters in Fig.~\ref{fig:cluster_shape}.  We used the following cuts to identify photons:  $\ell<100$~mm, $w_{T}<12$ mm, and $z_{cl}-z_{front}<300$~mm.

\begin{figure*}
    \centering
    \includegraphics[width=\linewidth]{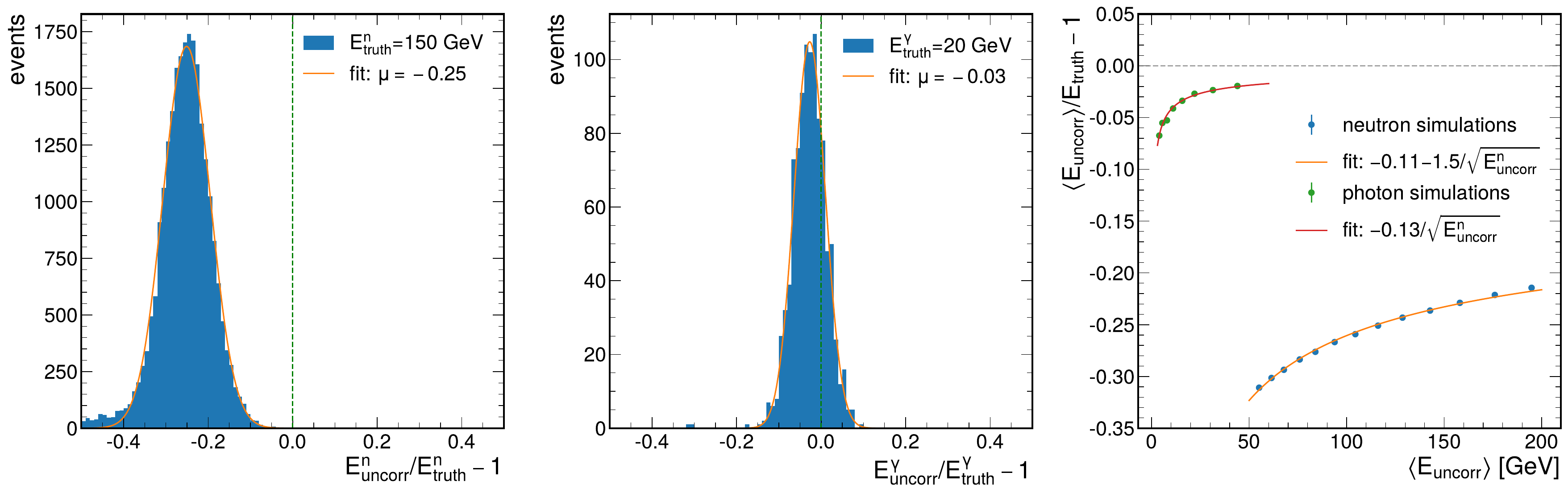}
    \caption{Left: relative difference between the uncorrected energy and the truth energy for neutrons.  Middle: same for single-photon events.  Right:  fits used for correcting the energies for neutrons and photons. }
    \label{fig:E_calib}
\end{figure*}

We selected events with two photon clusters and at least one other cluster\footnote{Hadronic showers in a high-granularity calorimeter like the ZDC~\cite{Milton:2024bqv} are expected to yield multiple clusters when using the topocluster algorithm.}.  The energy of each reconstructed photon, prior to corrections, is the energy of its associated cluster.  All clusters not associated with a photon were then associated with the neutron.  The first step to obtaining the neutron's energy is adding up the energies of all the non-photon clusters.  A correction factor is applied to the reconstructed neutron energy to account for the non-compensated nature of the ZDC~\footnote{This is a first-order correction, and more refined methods, such as software compensation~\cite{CALICE:2024imr} that weights each topocluster differently, could be applied to improve approach in future work.}:
\begin{equation}
E^n_{\rm corr}=E^n_{\rm uncorr}\frac{1}{1+A_n+B_n/\sqrt{E^n_{\rm uncorr}}},
\end{equation}
where the parameters $A_n$, and $B_n$ are $-$0.11 and $-$1.5, respectively, and $E^n_{\rm uncorr}$ is the neutron energy at the electromagnetic scale in GeV. These parameters were determined by taking the distribution of the $(E^n_{\rm uncorr}-E^n_{\rm truth})/E^n_{\rm truth}$ for single-neutron events at discrete $E^n_{\rm truth}$ values and then fitting them to Gaussian functions and taking the mean values for each fit.  An example is shown in the left plot of Fig.~\ref{fig:E_calib}.  We then determined the optimum values of $A$ and $B$ by fitting these mean values to the form $x_i/(1+A+B/\sqrt{x_i})$ where $x_i$ is the mean value of $E^n_{\rm uncorr}$ within the bin.  This fit is shown in the right panel of Fig.~\ref{fig:E_calib}.  

 We applied the same process to a sample of 8k single-photon events within the relevant energy range (5 to 60 GeV), and found that the photon energy correction could be expressed as
\begin{equation}
    E^{\gamma_i}_{\rm corr}=E^{\gamma_i}_{\rm uncorr}\frac{1}{1+A_\gamma+B_\gamma/\sqrt{E^\gamma_{\rm uncorr}}}, 
\end{equation}
where $A_\gamma=0$ and $B_\gamma=-0.13$.  An example fit for photons is shown in the middle panel of Fig.~\ref{fig:E_calib}, and the energy-dependence fit is shown in the right panel. As expected, we find that the amount of correction needed for photons is much smaller ($\approx2-6\%$) than for neutrons ($\approx20-30\%$).  It is only non-zero because the clustering algorithm misses a small amount of the electromagnetic shower.  Since the energy of a $\Lambda^0$ is dominated by the neutron, the correction for the photon energy has a small impact on the overall reconstruction of the $\Lambda^0$ energy.  

The energy of the $\Lambda^0$ is then the sum of those of the three particles:
\begin{equation}
    E^{\Lambda^0}_{\rm rec}=E^n_{\rm corr}+E^{\gamma_1}_{\rm corr}+E^{\gamma_2}_{\rm corr}.
\end{equation}

\subsection{Displaced Vertex and Mass Reconstruction}
A key challenge for determining the four-momenta of the three decay product particles is that the decay position of the $\Lambda^0$, $\vec x_{\rm vtx}$, is not known \textit{a priori}, and therefore the directions of these particles' momenta cannot be simply determined by the positions of the clusters.  However, we can make two assumptions about the decay position.  First, we assume the $\Lambda^0$ originated from the origin~\footnote{We neglect the finite beam-spot size expected at the EIC, as it introduces deviations from the nominal origin on the order of centimeters, whereas we are studying displaced vertices on the order of {$O(10)$m}.}, and therefore the decay position is along the line starting from the origin that has the same direction as the momentum of the $\Lambda^0$.  Secondly, the $\pi^0$ promptly decays at the same position where the $\Lambda^0$ decays, and the mass determined from the energies of the photons and the opening angles should match the value from the PDG, $\approx 135$~MeV~\cite{ParticleDataGroup:2024cfk}.  

We use the following iterative process, which we call IDOLA (Iterative Decay Origin for Lambda Analysis), to infer the decay position. In the first iteration, we assume that the decay took place at the origin and the momenta of the 3 particles is: 

\begin{equation}
\label{eqpi}
    \vec p_i=\sqrt{E_i^2-m_i^2}\left(\frac{\vec x^{\,\,i}_{\rm cl}-\vec x_{\rm vtx}}{\left|\vec x^{\,\,i}_{\rm cl}-\vec x_{\rm vtx}\right|}\right),
\end{equation}
where $\vec p_i$ is the momentum of the particle, $E_i$ is the energy, $m_i$ is the mass of the particle from the PDG (0 for photons, and $\approx 940$~MeV for neutrons~\cite{ParticleDataGroup:2024cfk}), $\vec x^{\,\,i}_{\rm cl}$ is the position of the cluster\footnote{For neutrons, which can have more than one cluster, this is the of position the highest-energy cluster associated with the neutron.}, and $\vec x_{\rm vtx}$ is the decay position in this iteration.  We then obtain from these momenta the momentum vector of the $\Lambda^0$ as the sum of those of the three particles,

\begin{equation}
\label{eqpLambda}
    \vec p_{\Lambda^0}=\vec p_n+\vec p_{\gamma_1}+\vec p_{\gamma_2}.
\end{equation}

The reconstructed mass of the $\pi^0$ is then
\begin{equation}
\label{eqmass}
    m^{\rm rec}_{\pi^0}=2\sqrt{E_{\gamma_1}E_{\gamma_2}}\sin\frac{\theta_{\rm open}}{2},
\end{equation}
where $\theta_{\rm open}$ is the opening angle between the two photons.   

In the second iteration, the decay position is halfway between the origin and the ZDC, along the direction of $\vec p_{\Lambda_0}$.  That is, 
\begin{equation}
    \label{eq:decay_pos}
    \vec x^{\,j}_{\rm vtx}=f^{(j)}\frac{\vec p_{\Lambda^0}}{\hat z\cdot \vec p_{\Lambda^0}} z_{\rm ZDC},
\end{equation}
where $\hat z$ is the direction of the proton axis, and $z_{\rm ZDC}=35.8$ m is the distance from the origin to the front face of the ZDC.  The superscript $j$ is the iteration number.  The value of $f^{(2)}$ is 0.5.  

\begin{figure}[h!]
    \centering
    \includegraphics[width=0.83\linewidth, trim=3cm 0 2.5cm 0,clip]{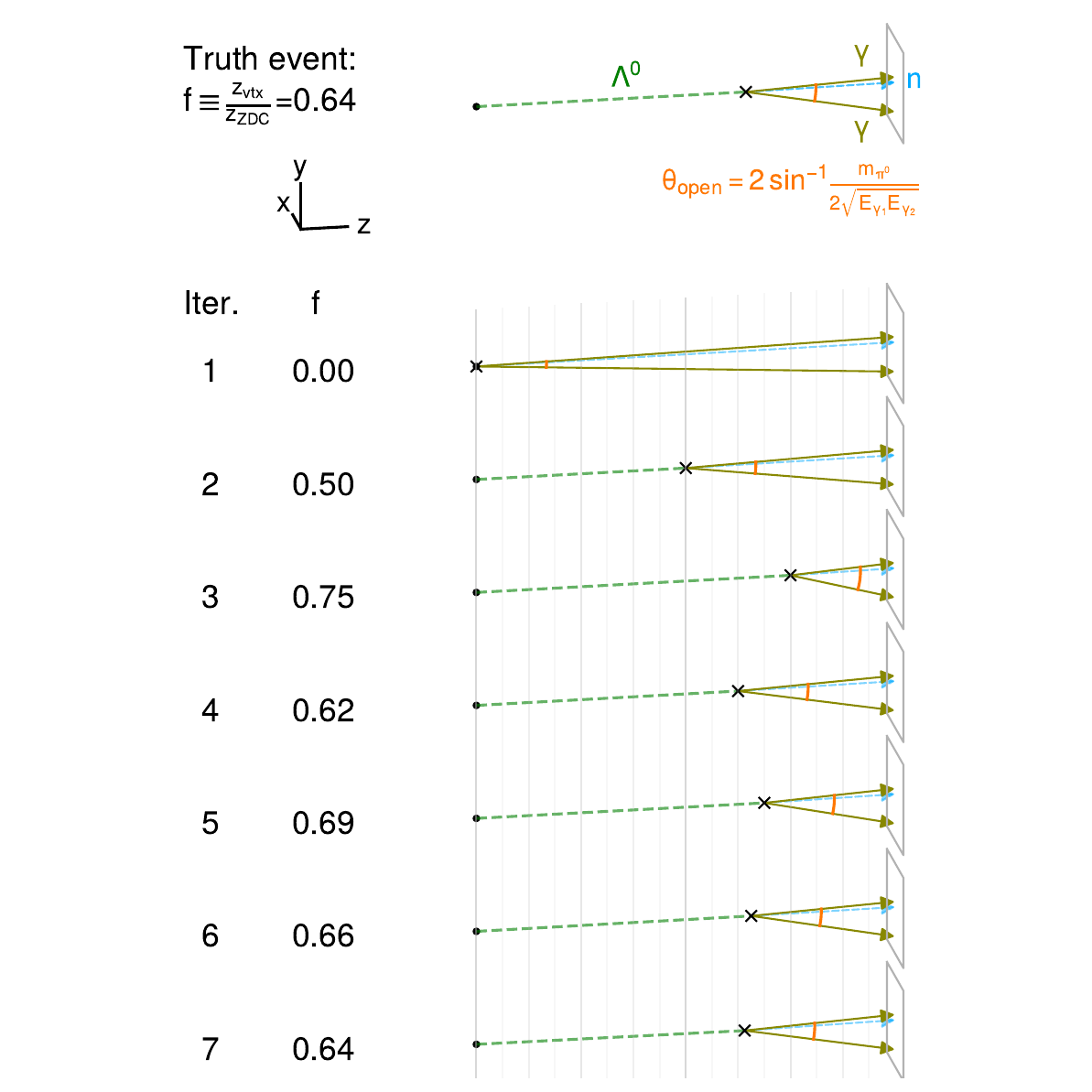}
    \caption{Illustration of the IDOLA algorithm.  In the top of the figure is the truth information for an event (not shown to scale). 
 In the first iteration of the algorithm, the reconstructed vertex is at the origin.  At the second iteration, it is halfway between the origin and the ZDC front face.  Since the angle between the two photons for this vertex-position hypothesis is too small, the hypothesis in the next iteration is closer to the ZDC.  The angle in the next iteration is too large, so the hypothesis after that is closer to the origin. }
    \label{fig:vertex_illustration}
\end{figure}
In subsequent iterations, equations~\ref{eqpi},\ref{eqpLambda}, and~\ref{eqmass} are recomputed. Equation~\ref{eq:decay_pos} is updated using 
\begin{equation}
    f^{(j)}=f^{(j-1)}\pm2^{-(j-1)}
\end{equation}
where $f^{(j-1)}$ is the value of $f$ from the previous iteration, and the sign $\pm$ is positive (negative) if $m^{\rm rec}_{\pi^0}$ from the previous iteration is less than (greater than) the PDG value.  For example, in the third iteration, the value of $f$ will either be 1/4 or 3/4, and in the fourth iteration, the value of $f$ will be an odd multiple of 1/8, \textit{etc}.  This process continues for 10 iterations, which has been found to be sufficient for convergence. The IDOLA algorithm is illustrated in Fig.~\ref{fig:vertex_illustration}.

\begin{figure*}[h!]
    \centering
    \includegraphics[width=\linewidth]{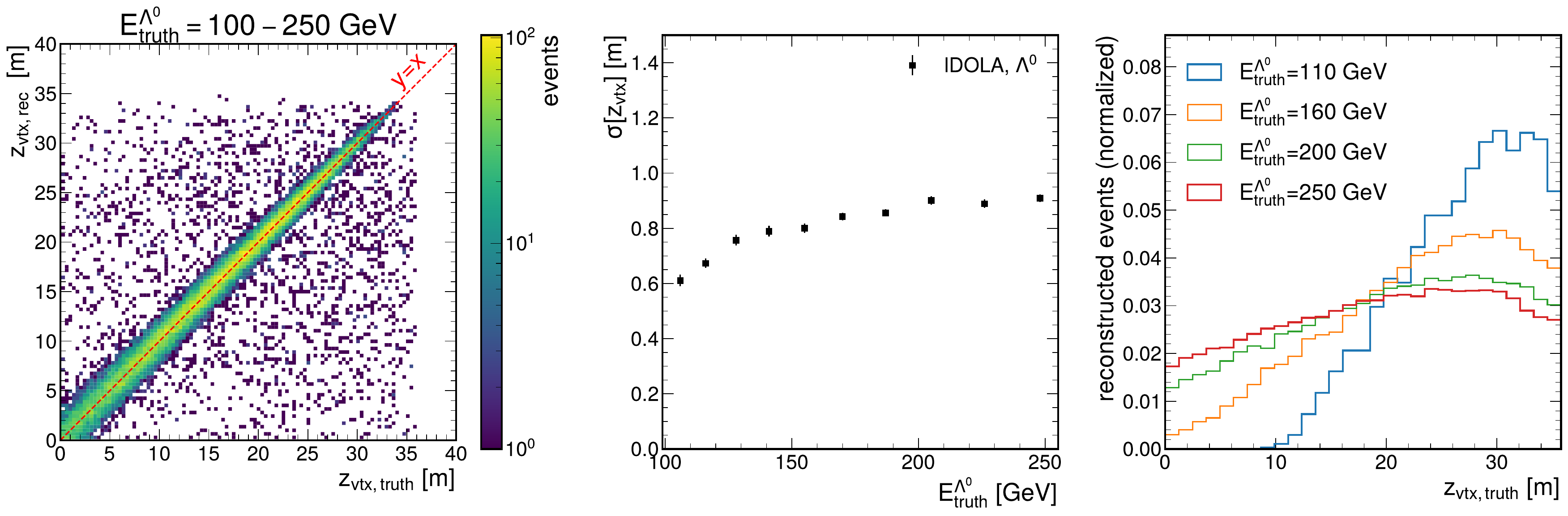}
    \caption{Left: truth decay position of $\Lambda^0$ ($x$ axis), compared to the reconstructed values ($y$ axis).  A thin red dashed line is shown at $y=x$.  Middle: the decay position resolution as a function of the $\Lambda^0$ energy.  Right: distribution for the decay positions of reconstructed events for different $\Lambda^0$ energies.  }
    \label{fig:vertex}
\end{figure*}
The left plot of Fig.~\ref{fig:vertex} shows the truth versus reconstructed values of $z_{\rm vtx}$ using IDOLA. We find that this method predicts $z_{\rm vtx}$ with negligible bias and primarily a Gaussian response.  The middle panel of Fig~\ref{fig:vertex} illustrates the $z_{\rm vtx}$ resolution, which ranges from approximately 0.6 to 0.9~m, depending on the energy.  

Projections of the true $z_{\rm vtx}$ values for reconstructed events are shown in the right plot of Fig.~\ref{fig:vertex}. The lowest-energy $\Lambda^0$s (100 GeV, shown in blue) are only reconstructed if the vertex is at least 10 m downstream of the origin, with the peak of the distribution occurring just a few meters upstream of the ZDC.  This is a consequence of geometric acceptance. At 250 GeV, the distribution of the decay positions for the accepted $\Lambda^0$s is closer to uniform.

\begin{figure*}[h!]
    \centering
    \includegraphics[width=\linewidth]{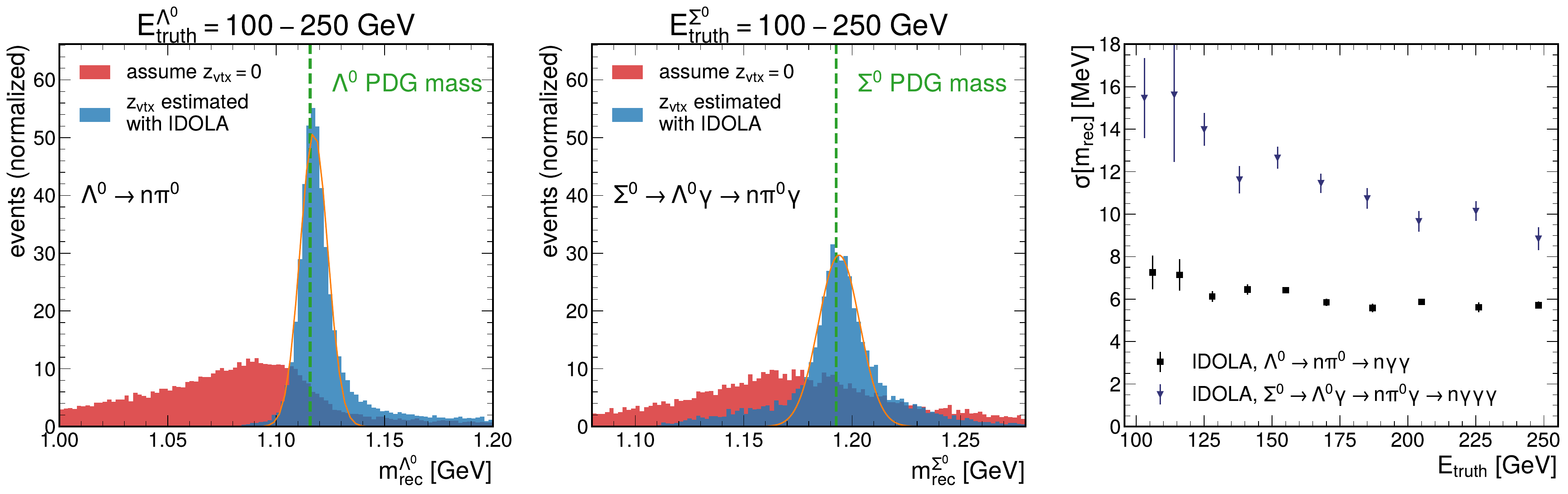}
    \caption{Left:  distribution for the reconstructed $\Lambda^0$ mass before (red) and after (blue) fitting the vertex position with the IDOLA algorithm.  Middle: Same for the reconstructed $\Sigma^0$ mass.  Right: the $\Lambda^0$ ($\Sigma^0$) mass resolution as a function of the $\Lambda^0$ ($\Sigma^0$) energy.}
    \label{fig:mass_res}
\end{figure*}

For further validation, we checked that the reconstructed values of the mass of the $\Lambda^0$ were consistent with the PDG value, $\approx 1116$~MeV~\cite{ParticleDataGroup:2024cfk}.  We show the distribution of the reconstructed $\Lambda^0$ masses in the left panel of Fig.~\ref{fig:mass_res}, and found that this is indeed the case.  

We repeated this exercise with simulated $\Sigma^0\rightarrow\Lambda^0\gamma\rightarrow n\pi^0\gamma\rightarrow n\gamma\gamma\gamma$ events\footnote{Studies of DEMP $\Sigma^0$ will cross-check extractions of kaon form factors~\cite{Arrington:2021biu}.}, where one of the photons comes from the prompt $\Sigma^0$ decay and the other two come from the $\pi^0$ decay. The events were selected as those with three clusters identified as being from photons, and at least one other cluster.  We tested all three hypotheses as to which pair of photons came from the $\pi^0$, and used whichever hypothesis produced the reconstructed $\Lambda^0$ mass closest to the PDG value, and kept only the events where the reconstructed masses of the $\Lambda^0$ for the best hypothesis were within 30 MeV of the PDG value.  Of the 630k $\Sigma^0$ events, only 13k events (2.1\%) passed these cuts. The resulting distribution is shown in the middle panel of Fig.~\ref{fig:mass_res}.  This distribution has a core centered at the PDG value ($\approx 1193$~MeV~\cite{ParticleDataGroup:2024cfk}) with tails on both sides.  

We performed fits to these distributions in various slices of the $\Lambda^0$ (or $\Sigma^0$) energy, and show the results in the right panel of Fig.~\ref{fig:mass_res}.  We find that the mass resolution for $\Lambda^0$ is around 6-7~MeV, while that of $\Sigma^0$ is about 8-16~MeV.  

\subsection{Polarization measurements}
The polarization of an ensemble of $\Lambda^0$ events can be determined using the distribution of the neutron direction in the center-of-mass (cm) frame of $\Lambda^0$. The probability distribution of the neutron direction in the cm frame is given by 
\begin{equation}
    \frac{dP}{d\Omega_n}=1+\alpha\vec\mathcal{P}^{\Lambda^0}_{\rm cm}\cdot\hat p^n_{\rm cm},
\end{equation}
where $\alpha$ is the analyzing power ($\approx$75\%), $\mathcal{P}^{\Lambda^0}_{\rm cm}$ is the $\Lambda^0$ polarization in the cm frame, and $\hat p^n_{\rm cm}$ is the direction of the neutron in the cm frame. Measurements of this polarization could be used in various spin-related studies as suggested in Ref.~\cite{Tu:2023few,Ji:2023cdh}, similar to previous measurements~\cite{PhysRevD.91.032004,ABT2006415,PhysRevD.40.3557,Kang:2021kpt}.

\begin{figure}
    \centering
    \includegraphics[width=\linewidth]{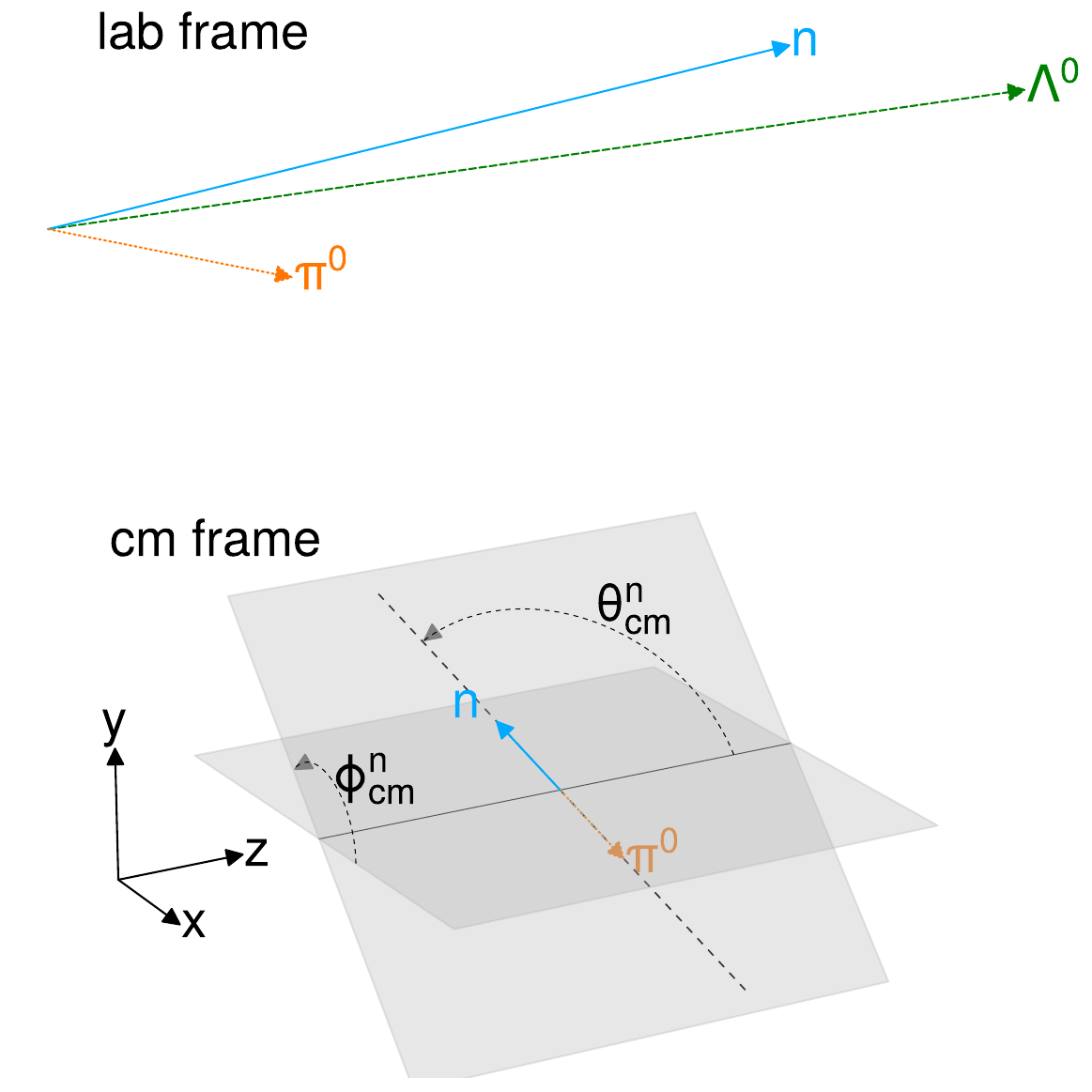}
    \caption{Illustration of the variables associated with the polarization of the $\Lambda^0$.}
    \label{fig:polarization}
\end{figure}

To obtain the cm direction of the neutron, we take the lab-frame direction, and boost it in the direction opposite that of the $\Lambda^0$, as follows:
\begin{equation}
    \left(\begin{array}{c}
         \vec p^{\,n}_{\rm cm}\\
         E^{\,n}_{\rm cm}
    \end{array}\right)=
    \left(\begin{array}{c c}
         \mathbf{I}_3+\frac{\gamma-1}{\beta^2}\vec \beta\,\vec\beta^{\,\,T} & -\gamma\vec\beta \\
         -\gamma\vec\beta^{\,\,T} & \gamma
    \end{array}\right)
    \left(\begin{array}{c}
         \vec p^{\,n}_{\rm lab}\\
         E^{\,n}_{\rm lab}
    \end{array}\right),
\end{equation}
where $\vec\beta$ is the velocity of the $\Lambda^0$ divided by the speed of light, and $\gamma=\frac{1}{\sqrt{1-\beta^2}}$.  From this, we define the angles of the neutron direction in the cm frame as 
\begin{equation}
    \label{eq:theta_cm}
    \theta^{\,n}_{\rm cm}={\rm atan2}\left(\sqrt{(p^n_{{\rm cm},x})^2+(p^n_{{\rm cm},y})^2}\, ,\,\,p^n_{{\rm cm},z}\right),
\end{equation}
and 
\begin{equation}
    \label{eq:phi_cm}
    \phi^n_{\rm cm}={\rm atan2}\left(p^n_{{\rm cm},y},p^n_{{\rm cm},x}\right).
\end{equation}
These are illustrated in Fig.~\ref{fig:polarization}.

\subsection{Graph Neural Networks}
\label{sec:AI}
We additionally performed a reconstruction using graph neural networks (GNNs) following the approach of Refs.~\cite{ATL-PHYS-PUB-2022-040, codesign,Acosta:2023zik, Milton:2024bqv}. This was done using the Graph Nets~\cite{GraphNets} library in \textsc{TensorFlow} (version 2.13.0)~\cite{tensorflow2015-whitepaper}.

The GNN was trained using the continuous data, with 25\% of the dataset used as a validation set, and tested with the discrete data. Both datasets are described in Section~\ref{sec:simulation}. We trained the GNN on continuous energies to prevent it from memorizing the discrete energies used in the test dataset.

For each event, the ZDC's hits were represented as a graph. Each node of the graph contained the energy and position of a reconstructed hit. The nodes were connected using a set of edges, which connect the information of the nearest $k=10$ neighboring hits defined with the k-nearest neighbors algorithm implemented in scikit-learn~\cite{scikit-learn}. We also added a global node containing the sum of the hit energies divided by the ZDC's sampling fraction to help the model learn the energy scale of the events. We normalized all the energies and positions in the nodes using a $z$-score normalization with means and standard deviations obtained from a subset of the data. The means and standard deviations were calculated separately for the training and test datasets.

We trained a dense neural network composed of four dense layers with 64 nodes each to predict the generated energy, polar angle, and azimuthal angles of the $\Lambda^0$ --- $E_{\rm truth},~\theta_{\rm truth}$, and $\phi_{\rm truth}$, respectively. We used the mean absolute error (MAE) for the loss function with equal weight given to each of the three regression targets. Each dense layer used the Rectified Linear Unit (ReLU) activation function~\cite{relu} and He-normal initialization \cite{HeUniform}. We trained the model with a batch size of 256 events using the Adam optimizer~\cite{adam} for 15 epochs, at which point the loss converged. We initialized the learning rate to $1e^{-3}$ and halved it every 5 epochs to a minimum of $1e^{-6}$.

In Sec.~\ref{subsec:lambda_identification}, we extended the model to do event classification. We trained the model using an equal mixture of single $\Lambda^0$ and single neutron events. The extended model predicts the $E_{\rm truth},~\theta_{\rm truth}$, and $\phi_{\rm truth}$ of the incident particle --- either the $\Lambda^0$ or the neutron --- as well as the particle type. We converted the model's output for the particle type to a probability that the incident particle was a $\Lambda^0$ using the sigmoid function. For training this combined regression and classification model, we used the modified loss function from Refs.~\cite{ATL-PHYS-PUB-2022-040, Milton:2024bqv}:
\begin{equation}
    \Lb= (1-\alpha)\Lb_{\rm classification} + \alpha \Lb_{\rm regression},
    \label{eqn:loss}
\end{equation}
where $\Lb_{\rm classification}$ is the binary cross-entropy loss, $\Lb_{\rm regression}$ is the mean absolute error loss, and $\alpha$ is a hyperparameter specifying the importance of classification versus regression. $\alpha$ was set to 0.75 for these studies, as was done in Refs.~\cite{ATL-PHYS-PUB-2022-040, Milton:2024bqv}. Energy, $\theta$, and $\phi$ were given equal weights in $\Lb_{\rm regression}$. We trained this network for 100 epochs as the loss converged more slowly compared to the $\Lambda^0$ regression model due to the increased amount of data and added complexity of classification.

We converted the probability that an event is a $\Lambda^0$ to a particle type by employing a probability cut: a ``loose" cut at 0.2, a ``medium" cut at 0.8, and a ``tight" cut at 0.98, which result in high, medium, and low efficiency, respectively, for classifying events as $\Lambda^0$. The efficiencies will be presented in Section~\ref{subsec:lambda_identification}.

For the polarization reconstruction in Section~\ref{subsec:polarization}, we used a GNN to predict $E_{\rm truth},~\theta_{\rm cm}$, and $\phi_{\rm cm}$ of the $\Lambda^0$. For this 3D regression, we again trained for 15 epochs only using the $\Lambda^0$ data and the MAE loss function with equal weights given to each of the regression targets.
\section{Performance}
\label{sec:performance}

\subsection{Lambda identification}
\label{subsec:lambda_identification}
In the conventional reconstruction, based on IDOLA, we selected $\Lambda^{0}$ candidates by requiring exactly two clusters that pass the photon-selection cuts (see Sec.~\ref{sec:conventional_recon}), at least one additional cluster in the event, and that the reconstructed mass of the $\Lambda^0$ lies within 30 MeV of the PDG value. These conditions define our signal events for the conventional reconstruction throughout the remainder of this paper. In the GNN-based reconstruction, we use the GNN classifier described in Sec.~\ref{sec:AI}.

We define the $\Lambda^{0}$ efficiency of a classifier as the ratio of the number of events identified as $\Lambda^0$, divided by the total number of events within the geometrical acceptance defined in Sec.~\ref{sec:reconstruction}.

\begin{figure}
    \centering
    \includegraphics[width=\linewidth]{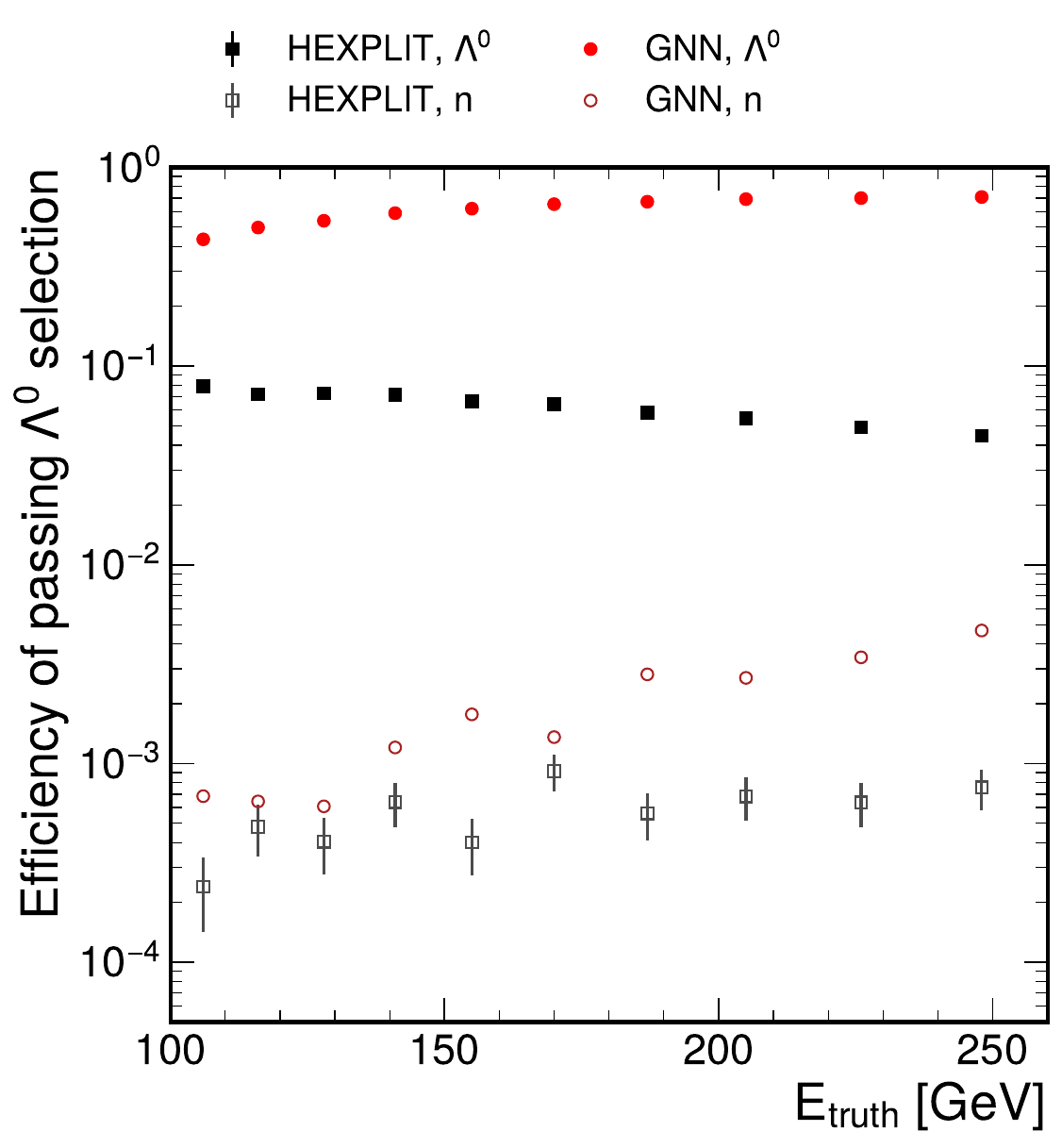}
    \caption{
    Efficiency of events passing the conventional (black squares) and GNN-based (red circles) $\Lambda^0$ selection.   The values for $\Lambda^0$ events are shown as filled symbols, while those for the neutron events are shown as open symbols.  
    }
    \label{fig:identification}
\end{figure}

Figure~\ref{fig:identification} shows the efficiency of reconstructing a $\Lambda^0$ using both the conventional reconstruction and the GNN classifier. We find that 4\%-8\% of the $\Lambda^0$ events pass the conventional reconstruction, whereas 43\%-70\% are correctly identified as $\Lambda^0$ events by the GNN. We also tested the same classifiers on single-neutron events and found that less than 1\% of the neutron events were misidentified by either classifier.

The GNN results presented here use the ``tight'' cut of 0.98 to convert the probability of an event being a $\Lambda^0$ into a particle classification. This cut provides the most similar level of neutron background rejection as the conventional approach while still achieving higher $\Lambda^0$ efficiency. The ``medium'' cut of 0.8 yields $\Lambda^0$ efficiencies between 61\% and 87\% with neutron backgrounds up to 3\%, whereas the ``loose'' cut of 0.2 achieves $\Lambda^0$ efficiencies of 83\%-99\% but with neutron backgrounds as high as 31\%.

The required rejection of neutrons will depend on the relative number of expected neutrons compared to $\Lambda^0$s for a given kinematic range of interest.  For reference, the cross sections for leading $\Lambda^0$ baryons at EIC energies were calculated in Ref.~\cite{Carvalho:2023kfb}, and were found to be around 1/3 of those of the neutrons.  In such a case, both the IDOLA-based method and the tight version of the GNN method would be sufficient, with the latter giving a better efficiency.    

Other sources of backgrounds, such as non-DEMP events, have not been studied, but those can be reduced with missing mass cuts to negligible levels~\cite{EICYR,Bylinkin:2022rxd}. Background from beam-gas interactions, or SiPM noise, could affect these numbers as they are not included. This can be studied in future refined studies.

At 100 GeV and $p_T$=0 GeV, the combined acceptance and efficiency is 2\%$\times$9\% or 0.018\% for conventional methods and 2\%$\times$40\%, or 0.8\% for the GNN method. These values increase to around 35\%$\times$5\%$\approx$1.7\% for conventional and 35\%$\times$70\%$\approx$24\% at 250 GeV. Assuming an integrated luminosity of 10 fb$^{-1}$, a cross-section of roughly 100~nb\footnote{Which is consistent with estimates of the $\Lambda^0$ production cross section from the DEMPgen event generator~\cite{Ahmed:2024grm} for the EIC at the 18 $\times$ 275 GeV setting.}, a $\Lambda^0\rightarrow n\pi^0$ branching ratio of 36\%~\cite{ParticleDataGroup:2024cfk}, and that the combined acceptance $\times$ efficiency averaged over the $\Lambda^0$ energy and $p_T$ range is $\approx$10\%, then we would expect a sample size of around 36k events.  

\subsection{Energy resolution}
\begin{figure}
    \centering
    \includegraphics[width=\linewidth]{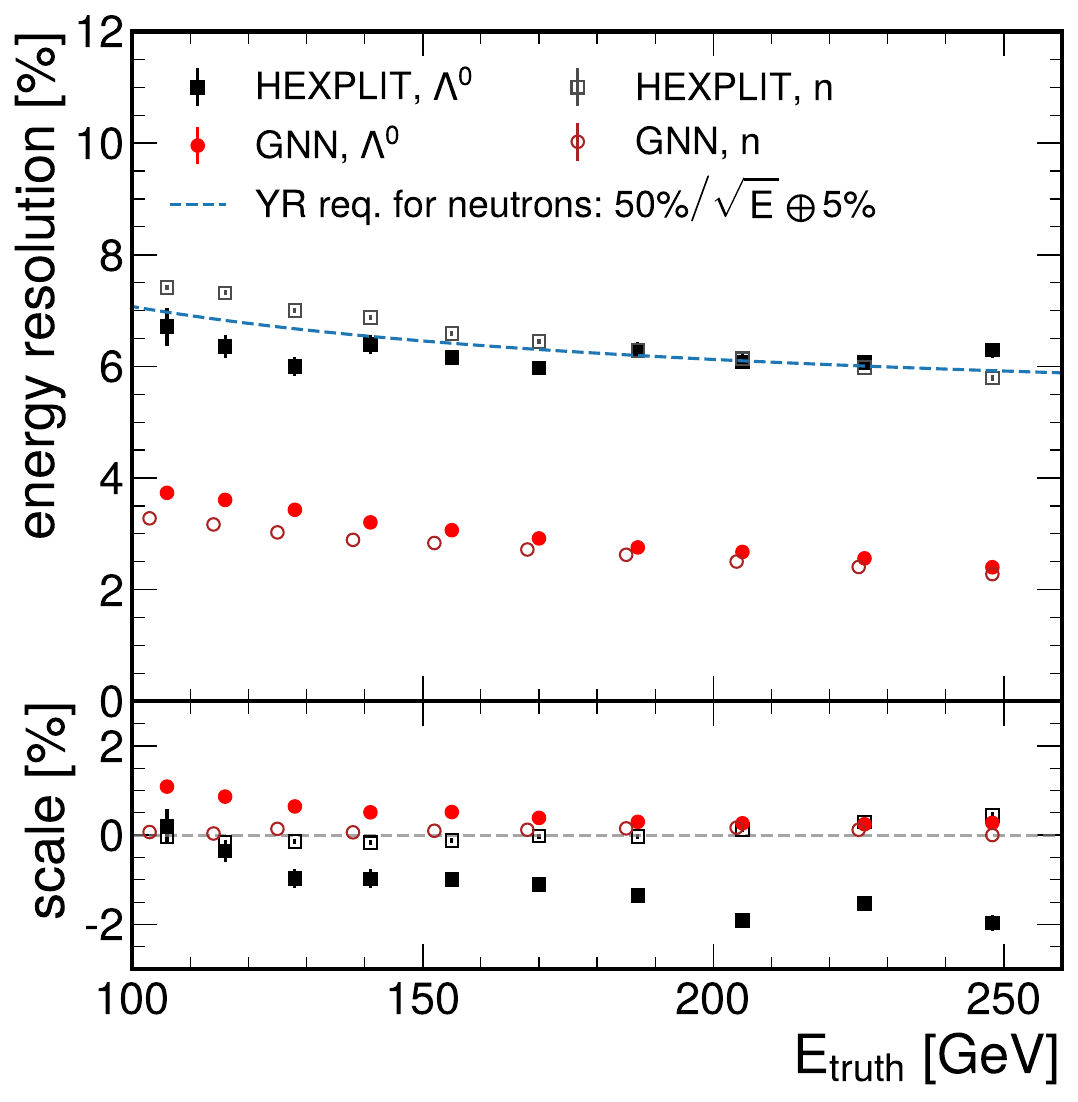}
    \caption{Energy resolution as a function of the $\Lambda^0$ energy.  For comparison, the values obtained in Ref.~\cite{Milton:2024bqv} for single neutrons are shown in gray.}
    \label{fig:E_res}
\end{figure}

To determine the resolution of the energy reconstruction for both methods, we took the distribution of $(E_{\rm rec}^{\Lambda^0}-E_{\rm truth}^{\Lambda^0})/E_{\rm truth}^{\Lambda^0}$, and fit these to Gaussian functions.  The resolutions are the standard deviations of these fits and the energy scales are the means of these distributions.  We show the resolutions for both methods in Fig.~\ref{fig:E_res}. 
 We found that the energy scales are within $\approx$2\% of unity for both methods.  The GNN outperforms the conventional method, and achieves an energy resolution of $38\%/\sqrt{E}$, which is similar to the values obtained for single-neutron events with a GNN in Ref.~\cite{Milton:2024bqv}.  
 
 We also compare our conventional reconstruction to a similar conventional reconstruction of single neutrons.  Here the same HEXPLIT and topoclustering algorithms are used as in the $\Lambda^0$ case, and the neutron energy is the sum of the cluster energies.  This is considerably worse than the values cited in Ref.~\cite{Milton:2024bqv} for the ``strawman'' algorithm (which added the energies of all reconstructed hits), since part of the shower energy is not included in the clusters.  This indicates that the clustering algorithm could benefit from further adjustments and fine tuning, to allow for better inclusion of hits within the clusters while avoiding merging showers from different incident particles into the same cluster.   
 
 We also compare this to the EIC YR's requirement for neutron reconstruction, $50/\sqrt{E}\%\oplus 5\%$~\cite{EICYR}. We find that the conventional reconstruction for neutrons straddles this curve and for $\Lambda^0$s it is just below the curve.  However, the resolutions with the GNN for both types of particles are well below this curve.

\subsection{Angular resolution}
\label{sec:angularresolution}
\begin{figure}
    \centering
    \includegraphics[width=\linewidth]{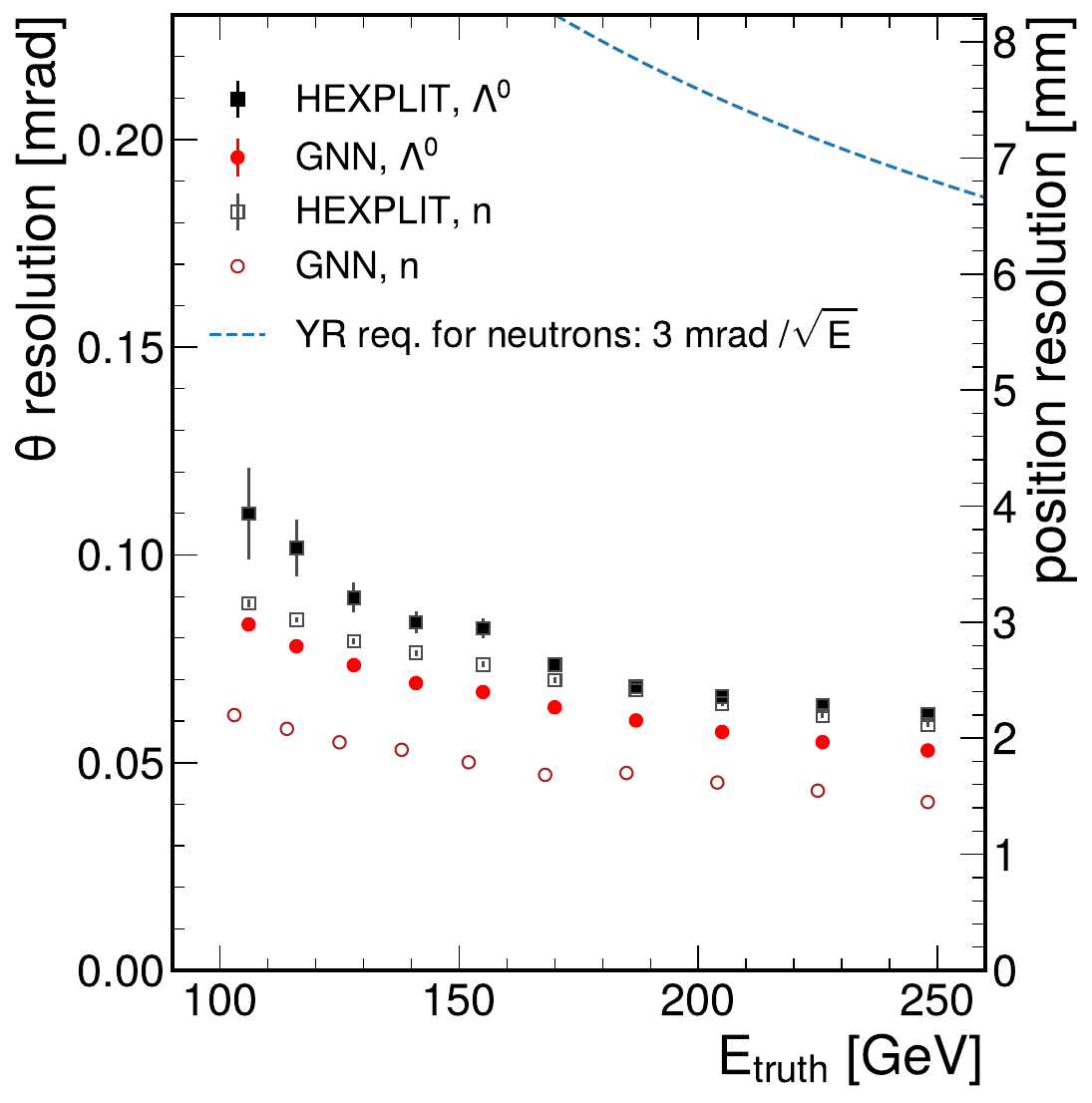}
    \caption{Polar angle resolution as a function of the $\Lambda^0$ energy.  For comparison, the values obtained in Ref.~\cite{Milton:2024bqv} for single neutrons are shown in gray.}
    \label{fig:theta_res}
\end{figure}

We determined the distributions of $\theta^{\,\Lambda^0}_{\rm rec}-\theta^{\,\Lambda^0}_{\rm truth}$, and fit these to the Gaussian functions.  The resolutions in $\theta$ were then the standard deviations of these fits, which we show in Fig.~\ref{fig:theta_res}. The resolutions for both algorithms were only slightly worse than those obtained for single neutrons\footnote{for the GNN, we compare to the results from Ref.~\cite{Milton:2024bqv}.  For the conventional reconstruction, we compare to a similar conventional reconstruction of neutrons.}.  However, the resolutions for both algorithms for either type of particle are much better than the requirement specified for neutrons in the EIC YR~\cite{EICYR} (3 mrad$/\sqrt{E}$).  

\subsection{Transverse momentum resolution}
\begin{figure}[h!]
    \centering
    \includegraphics[width=\linewidth]{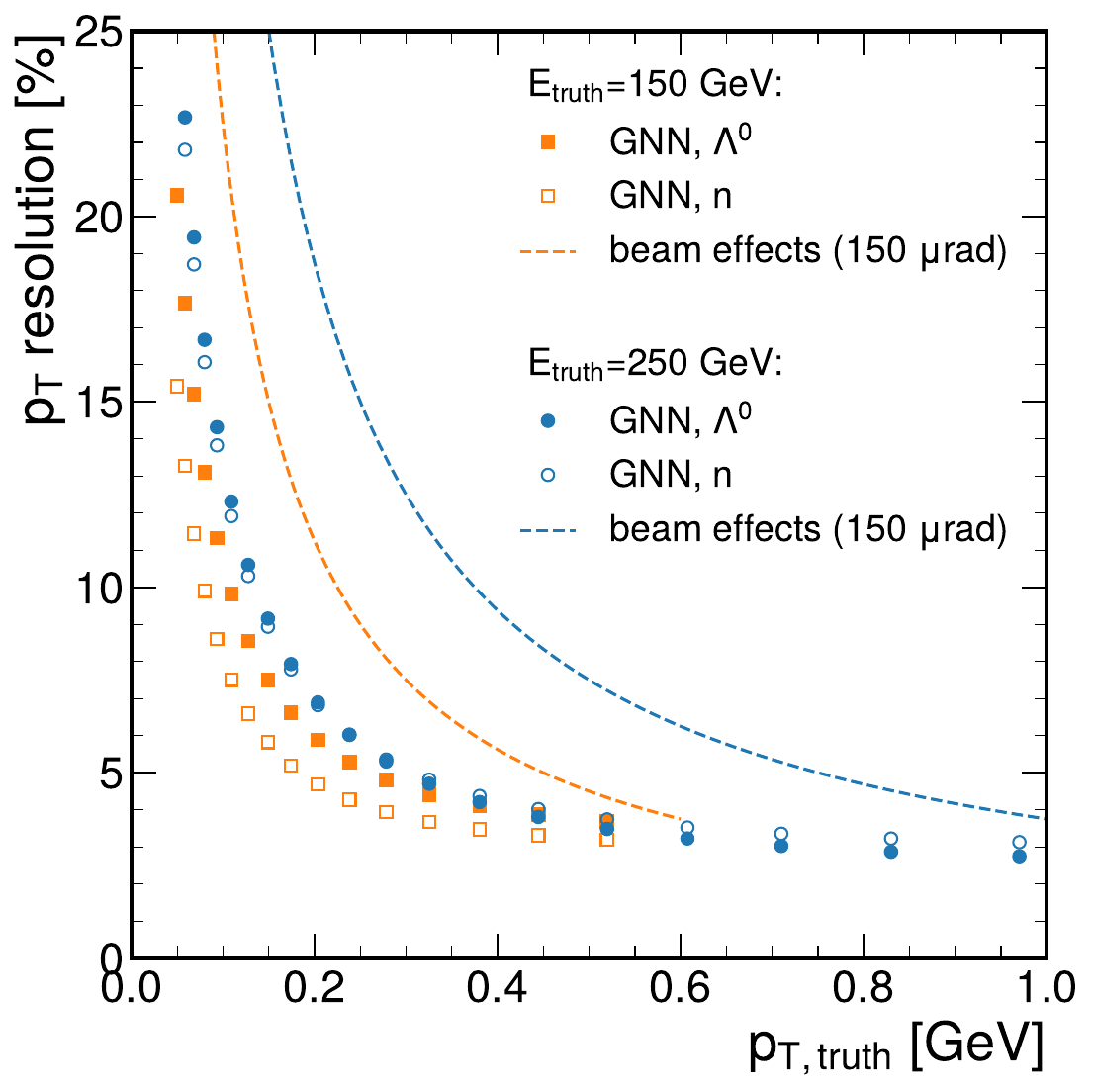}
    \caption{The resolution for reconstructing the transverse momentum of $\Lambda^0$ (filled symbols) and neutrons (open symbols) using a GNN at 150 GeV (blue) and 250 GeV (orange).  These values reflect the resolution without beam effects.  The beam effects terms for $\sigma_{\rm beam}$=150 $\mu$rad are shown as dashed curves.}
    \label{fig:ptres}
\end{figure}

Since the transverse momentum is approximately $p_T\approx E\theta$, the resolution can be approximated by
\begin{equation}
    \frac{\Delta p_T}{p_T}\approx \frac{\Delta E}{E} \oplus \frac{E\,\Delta\theta}{p_T}.
    \label{eq:ptres}
\end{equation}
We calculated the $p_T$ resolution by plugging the energy and polar-angle resolution values with the GNN reconstruction determined in the previous two sections into Eq.~\ref{eq:ptres}.  We show our results in Fig.~\ref{fig:ptres}, and compare this to the results for single neutrons with a GNN in Ref.~\cite{Milton:2024bqv}.

However, due to the spread of the proton beam, $\sigma_{\rm beam}=\mathcal{O}$(100~$\mu$rad), the actual $p_T$ resolution is worse than the ZDC resolution.  With beam effects, the expression for the $p_T$ resolution becomes:
\begin{equation}
    \frac{\Delta p_T}{p_T}\approx\frac{\Delta E}{E} \oplus \frac{E\,\Delta\theta}{p_T}\oplus \frac{E\,\sigma_{\rm beam}}{p_T}.
    \label{eq:ptres_beam}
\end{equation}
  We show the term from the beam effects as dashed curves in Fig.~\ref{fig:ptres} evaluated at $\sigma_{\rm beam}$=150 $\mu$rad, which is the value given in the YR for the high-luminosity configuration at 18 $\times$ 275~GeV~\cite{EICYR}. We found that beam effects will be the driver for $p_T$ resolution for both $\Lambda_0$ and neutrons.

\subsection{Polarization reconstruction}
\label{subsec:polarization}
\begin{figure}[h!]
    \centering
    \includegraphics[width=\linewidth]{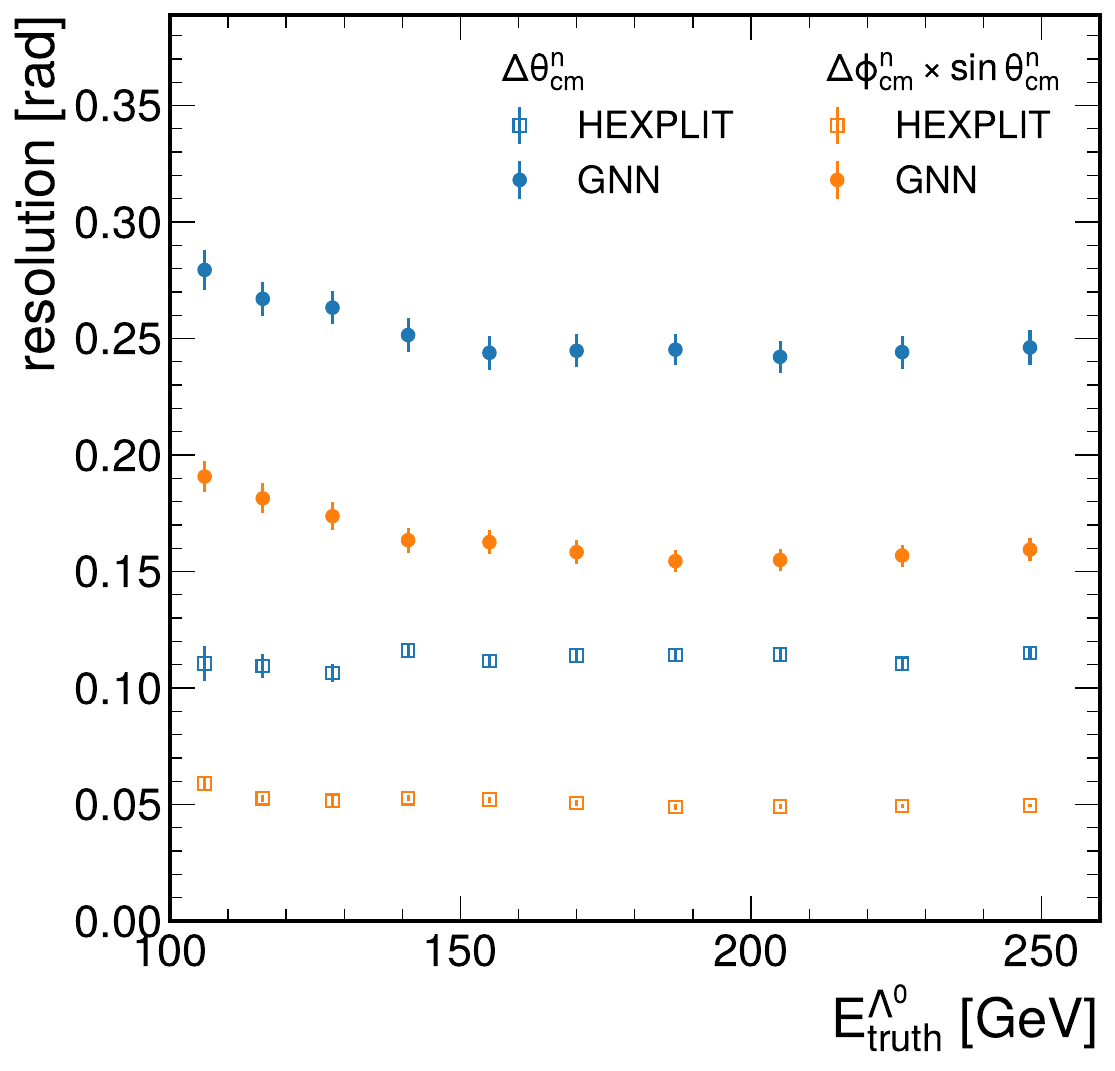}
    \caption{Resolution of the direction of the neutron in the rest frame of the $\Lambda^0$.  The polar (azimuthal) resolution is shown as blue circles, (orange squares).  We compare the results from the conventional reconstruction, based on the HEXPLIT~\cite{hexplit} and IDOLA algorithms, (open symbols) with those from the GNN reconstruction (filled symbols).}
    \label{fig:neutron_direction}
\end{figure}

We compared the reconstructed values of $\theta^{\,n}_{\rm cm}$ and $\phi^{\,n}_{\rm cm}$ to the truth values, and determined the resolutions using Gaussian fits to the distributions of the residuals $\theta^{\,n}_{\rm cm,rec}-\theta^{\,n}_{\rm cm,truth}$ and $(\phi^{\,n}_{\rm cm,rec}-\phi^{\,n}_{\rm cm,truth})\times\sin\theta^{\,n}_{\rm cm, truth}$.  The results are shown in Fig.~\ref{fig:neutron_direction}.  The polar angular resolution with the conventional reconstruction was determined to be around 100$-$120 mrad, while the azimuthal angle resolution was determined to be around 50$-$60~mrad.  We likewise calculated the resolutions for these variables with the GNN reconstruction, and found that the conventional algorithm outperforms the GNN reconstruction for both variables. We note that the conventional method requires at least three clusters, making it more selective in the events it accepts compared to the GNN reconstruction. This biases the conventional reconstruction toward events where the incident particles are more spread out. Since the measurement of the center-of-mass angles depends on the estimation of neutron energy, events with overlapping neutron and photon showers naturally lead to worse performance. This explains the better performance of the conventional method compared to the GNN in this case. This contrasts with the angle measurements discussed in Sec.~\ref{sec:angularresolution}, where only the overall $\Lambda^{0}$ momentum matters, not the neutron energy per se. 

\FloatBarrier
\section{Summary}
\label{sec:summary}

We have conducted a feasibility study for reconstructing high-energy $\Lambda^0$ baryons at the EIC through their neutral decay channel, $\Lambda^0 \to n\pi^{0}$, using a high-granularity Zero Degree Calorimeter~\cite{Milton:2024bqv}. The main challenge with this channel is estimating the $\Lambda^0$ displaced vertex, which, at the relevant EIC energies, is expected to range from meters to tens of meters---potentially the longest distances ever measured for displaced vertices in any experiment. We addressed this challenge using an iterative method that explicitly estimates the displaced vertex, as well as an AI-based approach that resolves it implicitly.

The $\Lambda^0$ energy and polar angle resolutions are comparable to those obtained for single neutrons in previous studies~\cite{Milton:2024bqv}, indicating that the $\Lambda^0 \to n\pi^{0}$ performance is primarily driven by the neutron measurement. This result is partly due to the decay kinematics in the center-of-mass frame, where most of the energy is carried by the neutron, which is also more challenging to measure than the $\pi^{0}$. Additionally, we presented performance studies for measurements of $\Lambda^0$ polarization with the angular distibution of neutrons in the $\Lambda^0$ center-of-mass frame.

Our findings extend prior $\Lambda^0$ feasibility studies~\cite{Tu:2023few,Aguilar:2019teb,Arrington:2021biu}, which focused on low-beam energy configurations with the charged-decay channel $\Lambda^0 \to p\pi^-$. We argue that the $n\pi^0$ and $p\pi^-$ channels are complementary. Notably, unlike the $p\pi^-$ channel, the $n\pi^0$ channel achieves higher acceptance at higher energies, allowing access to lower values of $x$ and higher $Q^{2}$ for future meson-structure studies. 

Furthermore, we demonstrated that the $\Lambda^0 \to n\pi^{0}$ decay channel is feasible using the SiPM-on-tile ZDC design~\cite{Milton:2024bqv} as a stand-alone detector which provides the dual function of hadronic and electromagnetic calorimetry at the higher proton energy.    

While the performance we have estimated indicates that the ZDC could exceed the EIC Yellow Report requirements for neutron and $\Lambda^0$ measurements with a comfortable margin, more realistic simulations, including backgrounds such as noise from SiPMs, represent a natural avenue for future studies. These future studies will focus on investigating the resilience of the methods described here to noise.

The methods described here could also be adapted for lower energies at the EIC or for the lower energies expected at the EiCC in China. However, this would require detector subsystems with the same design but covering larger polar angles. For such cases, a design similar to the ePIC high-granularity calorimeter insert~\cite{Insert}, which covers the range $3 < \eta < 4$, could be extended to the region $4 < \eta < 5$ with a different beam-pipe design. Such an approach could work for the EiCC kinematics~\cite{Xie:2021ypc} and provide excellent position resolution.

This study highlights the pivotal role of the ZDC in exclusive reactions at the EIC. Additionally, it showcases the imaging capabilities of SiPM-on-tile calorimeter technology and the power of integrating it with advanced AI/ML methods.

\section*{Code Availability}
The code for data processing, model training, and model inference are found at \url{https://github.com/eiccodesign/regressiononly/tree/lambda_decay}.
The GNN models stored as \textsc{TensorFlow} models and in the \textsc{ONNX} format are available in~\cite{milton_2024_14518203}.  A stand-alone demo of the IDOLA algorithm, presented here, can be found in~\cite{paul_2024_14518550}.

 \section*{Acknowledgments}
We thank the ePIC Collaboration for their valuable feedback during the preliminary phases of this work. 

We acknowledge support from DOE grant award number DE-SC0025600.  We also acknowledge support by the MRPI program of the University of California Office of the President, award number 00010100. Ryan Milton was supported by a ``HEPCAT fellowship'' from DOE award DE-SC0022313.

\FloatBarrier
\renewcommand\refname{Bibliography}
\bibliographystyle{utphys} 
\bibliography{bibio.bib} 

\appendix
\end{document}